\documentclass[times,twocolumn,authoryear,final]{elsarticle}
\usepackage{jasr}
\usepackage[export]{adjustbox}
\usepackage{framed,multirow}
\usepackage{amssymb}
\usepackage{latexsym}
\usepackage{commath}
\usepackage{graphicx}
\usepackage[table,xcdraw]{xcolor}
\usepackage[switch]{lineno}
\usepackage{xcolor, soul}
\sethlcolor{green}
\usepackage{lipsum, babel}
\usepackage{url}
\usepackage{xcolor}
\usepackage{amsmath,array,graphicx}
\definecolor{newcolor}{rgb}{.8,.349,.1}
\usepackage{caption}
\usepackage{subcaption}
\usepackage{array, booktabs, caption}
\usepackage{makecell}
\usepackage{tikz}

\usepackage[citebordercolor=white]{hyperref}
\usepackage{tabularray}
\usepackage{float}
\usepackage{aas_macros}
\journal{Advances in Space Research}

\begin{document}
\verso{R.J.S Airey \textit{et. al.}}

\begin{frontmatter}

\title{A comprehensive survey of the GEO-belt using simultaneous four-colour observations with STING}
    \author[Warwick-astro,Warwick-csda]{Robert J S \snm{Airey}}\corref{cor}
    \cortext[cor]{Corresponding author}
    \ead{Robert.Airey@warwick.ac.uk}
    \address[Warwick-astro]{Department of Physics, University of Warwick, Coventry, CV4 7AL (UK)}
    \address[Warwick-csda]{Centre for Space Domain Awareness, University of Warwick, Coventry, CV4 7AL (UK)}
    \address[DSTL]{Defence Science and Technology Laboratory, Porton Down, Salisbury SP4 0JQ, UK}
    \author[Warwick-astro,Warwick-csda]{Paul \snm{Chote}}
    \author[Warwick-astro,Warwick-csda]{James A \snm{Blake}}
    \author[Warwick-astro,Warwick-csda]{Benjamin F \snm{Cooke}}
    \author[Warwick-astro,Warwick-csda]{James \snm{McCormac}}
    \author[Warwick-astro,Warwick-csda]{Phineas \snm{Allen}}
    \author[Warwick-astro,Warwick-csda,DSTL]{Alex \snm{MacManus}}
    \author[Warwick-astro,Warwick-csda]{Don \snm{Pollacco}}
    \author[Warwick-astro,Warwick-csda]{Billy \snm{Shrive}}
    \author[Warwick-astro,Warwick-csda]{Richard \snm{West}}
\received{}
\finalform{}
\accepted{}
\availableonline{}
\communicated{}
\begin{abstract}
%%%
Colour light curves of resident space objects (RSOs) encapsulate distinctive features that can offer insights into an object's structure and design, making them an invaluable tool for classification and characterisation. We present the results of the first large systematic colour survey of the GEO (Geostationary Earth Orbit) belt in which we obtain full-night multi-colour light curves for 112 active geostationary objects between April and May 2023. This survey was undertaken during the `glint' season using the STING telescope located at the Roque de Los Muchachos observatory in the Canary Islands.

Colour light curve maps were created to compare and contrast the colours between different satellites and bus configurations. We find that satellites with BSS-702 and STAR-2 buses can be effectively distinguished from the colour measurements on these maps, but comparing the average colour of individual satellites within given solar equatorial phase angle ranges shows that it is difficult to distinguish between bus configurations based on colour alone. We also find tentative evidence to suggest that there is a relationship between colour and time spent on orbit for the Eurostar-3000 class satellites, which is unseen behaviour within other bus configuration classes. The satellites in our sample exhibit `redder' colours than the Sun, which is in agreement with previous findings. 

We found common light curve features such as symmetrical colour changes as well as unique regions of short timescale glinting which are `bluer' than other regimes within the colour light curves. If these features are indeed seasonal, this would be a powerful characterisation tool.  We are able to detect and resolve features in the light curve of the LDPE-3A satellite related to manoeuvres being performed. 

%Finally, we measured the solar panel offsets of 54 satellites in our sample and found variation in the type of colour response with the majority not having any associated colour change change across the glints compared to them shifting towards `redder' or `bluer' colours. 
Finally, we measured the solar panel offsets of 54 satellites in our sample and found variation in the type of colour response. The majority of which did not exhibit any colour change across the solar panel glints compared to them shifting towards 'redder' or 'bluer' colours.

\end{abstract}


\end{frontmatter}

%% For linenumbersj

\section{Introduction}
Active satellites in GEO (Geostationary Earth Orbit) serve a variety of purposes which include; telecommunications, meteorology, data relaying and military activities, thus they are a very important part of our infrastructure. The risks posed to satellites in GEO are due to the satellites themselves becoming more manoeuvrable and existing debris in the regime. Therefore, it is necessary that these objects are actively monitored and characterised. Efforts towards a sustainable space environment have been outlined in works such as but not limited to; \cite{Blake_2022}, \cite{2022NatAs...6..428L}, \cite{2023arXiv231109504L} and \cite{2024AcAau.217..238B}.

Given that the GEO regime is approximately 36,000\,km above the Earth's surface, resolved imagery is infeasible from the ground and radar detection techniques are less sensitive. Therefore, ground-based photometric observations are required to characterise these objects. Observations of GEO satellites in multiple bands can be particularly useful in investigating characteristics such as; the effects of space-weathering, surface changes through analysis of colour index profiles and the identification of mounted components and their status.

There have been a range of studies that have constructed detailed observational strategies and data reduction pipelines to focus on characterising objects in the GEO regime. The common approach is analysis of optical light curves~\citep[see e.g.,][]{benson2017light,2021AdSpR..67..360B,chote2019precision,2021spde.confE..49K}.

Long-term observations of GEO objects can provide insight into orbital manoeuvres through anomalous long-term changes in brightness \citep{XiaoFen2021LongtermPS} or probing the phenomenon of satellite glints and expected brightness changes as a satellite spends more time in orbit \citep{Bold2018FireOPALAO}. 

Optical light curves can be used to identify and calculate how far a satellite's solar arrays are offset from the sun \citep{Payne2006SSAAO}. This is useful as these solar panel offsets have a direct implication on the status and health of a given GEO satellite and thus can be used to characterise them. The primary causes for this offset pointing can be: 
\begin{itemize}
    \item  Ensuring the satellite has the required electrical generation from launch until its designated end of operation, the offsets are likely to be implemented intentionally at the beginning of operations as panels would generate too much power
    \item Uneven degradation of the light sensors fitted on the solar panels can lead to unintended pointing when trying to balance the voltages across the panels \citep{Payne2006SSAAO}
    \item To balance the impact of the solar radiation pressure
\end{itemize}

Measured solar panel offset angles can also be used to estimate the albedo area product of a solar panel, which is useful as a concept to potentially infer the size of a given Resident Space Object (RSO) \citep{2019JAnSc..66..170P}. The aforementioned properties, collectively investigated, can be used in principle to construct a 3D photometric model to synthesise light curves and probe the relationship between the satellite design (e.g antenna reflectors, panel offsets, basic shape model) and the brightness changes observed in both the real and synthesised light curve \citep[e.g.][]{Skuljan2018PhotometricMO}.

There are limitations to such studies that implement observations across only one band. Observations taken across a single filter will most likely be unable to probe features relating to materials or specific components given the dependency of reflectance on wavelength, whereas observations taken across multiple filters likely allow for the identification and characterisation of surface materials and mounted components (solar panels, antennas etc).

However, colour measurements of observed GEO satellites are affected by complex factors such as the illumination geometry, variations in the spectral energy density of reflected light near specular peaks, and the differing BRDFs (Bidirectional Reflectance Distribution Functions) \citep{2012tres.book.....H} of satellite surface materials \citep{2023AdSpR..72.3802Z}. Generally, it is found that using colour solely for satellite discrimination is extremely difficult given the variability of colour across a single night of observations \citep{2015amos.confE..74J}.

Despite these complexities, there have been numerous studies that have used colour measurements of GEO satellites to investigate and constrain properties such as ageing, bus configuration and manufacturing markers. Overall, the key findings from these studies are that dry masses and bus configurations seem to significantly impact colour measurements \citep{2023sndd.confE..64D} with evidence of clustering around certain colours within colour-colour diagrams and a null correlation between colour and time spent on-orbit \citep{SCHMITT2020326}. Multi-colour photometry can also be useful in characterising and classifying tumbling space debris given the relationships between rotational phase and colour indices \citep{2022cosp...44.3160S}.

There are common limitations of the studies described above. We note that often these surveys utilise telescopes which have a narrow field of view, which impacts upon the ability to conduct a large survey at GEO, i.e., observing as many GEO targets of interest as we can within a field of observation~\citep{blake2023exploring}. Previous studies have also only focused on single object observations, which span over multiple months. Whilst this is indeed useful for characterising these objects individually and probing how seasonal and attitudinal variations impact light curve features, there are a very large number of satellites in the GEO regime and these vary in shape, size, design (usually in accordance with their bus configurations) and position on the sky. Therefore, it is going to be difficult to apply the observations of a singular satellite across a range of satellites without many assumptions being made. In addition, multi-colour photometric surveys of GEO satellites have often made use of non-simultaneous colour measurements, which introduce systematic errors when the target's intrinsic brightness changes on a similar timescale to the observations. Another common limitation is that colour indices are often quoted at particular phase angles, neglecting the fact that the colour can change throughout the night as different components of the target become illuminated.

A large number of objects that we observed within this survey were on inclined orbits, and observed only for a limited part of the night when they passed through the field of view. The majority of these objects are uncontrolled and perturbed by the gravitational influence of the Moon, the Sun and the oblateness of the Earth, which has the effect of causing the inclination of the orbits to vary between $\pm$ 15\,$^\circ$ on a $\approx$ 53 year period \citep{McKnight2015IAC15A673x27478OC}. This meant that those objects will oscillate along the north-south direction over a day.

The work we present here accounts for many of those limitations. This is the first large systematic colour survey of the GEO regime. We present full-night, multi-colour light curves for over 100 active geostationary satellites observed over a 5 week span. Section \ref{sec:instr-obs} introduces the instrumentation and characterisation of STING and our observation campaign. Sections \ref{sec:Results} and \ref{sec:Analysis} detail the demographics following our observation campaign and describe the analysis of our results. Section \ref{sec:discussion} and Section \ref{sec:summary} contextualise the findings from our results and discuss future work.
\begin{figure}[H]
    \centering
    \includegraphics{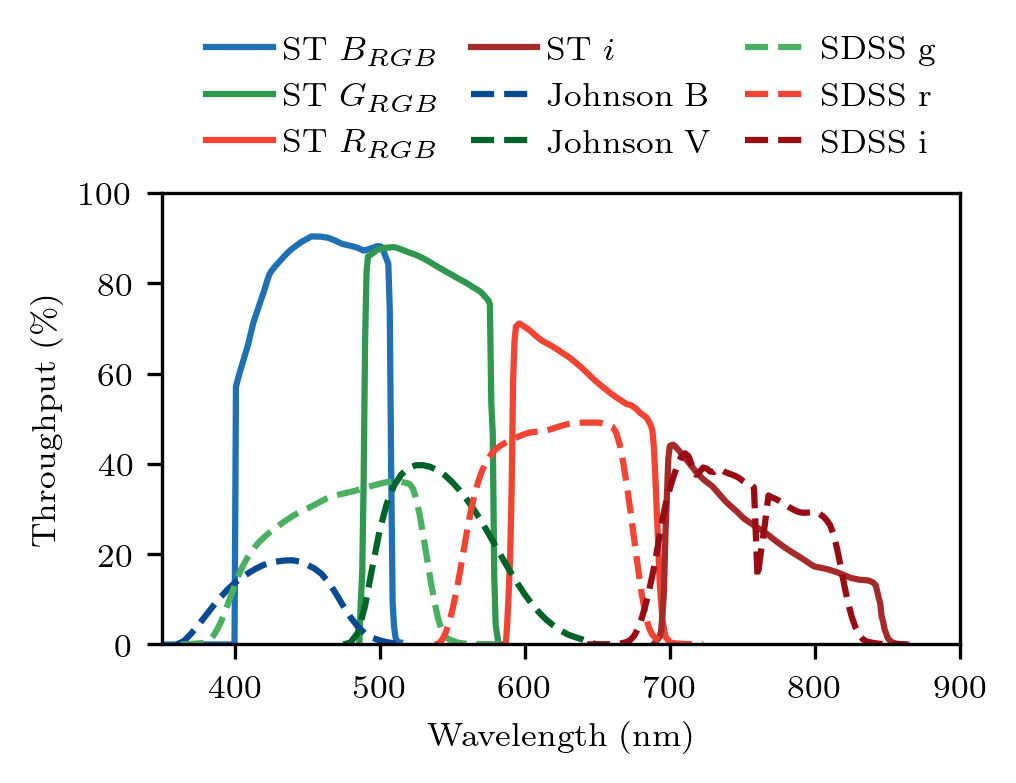}
    \caption{Total throughput as a function of wavelength for the cameras and filters used in STING (solid lines,  labelled by ST). For comparison, those used in the APASS AAVSO Photometric All Sky Survey (Henden et al., 2015) are shown (dashed lines).}
    \label{fig:Bandpass}
\end{figure}
\section{Instrumentation and Observation Campaign}\label{sec:instr-obs}
Observations for this survey were carried out using the STING telescope, part of the University of Warwick's facilities within the Roque de los Muchachos observatory on La Palma in the Canary Islands. STING was commissioned in late 2022 / early 2023 as a major refurbishment and upgrade of the SuperWASP-North exoplanet survey instrument.

\subsection{STING Hardware and Software}\label{subsec:SWASP-dets}
The original SuperWASP (Super Wide Angle Search for Planets) project \citep{Pollacco2006} was comprised of two observatories, SuperWASP-N on La Palma and SuperWASP-S in South Africa. The primary objective of SuperWASP was detecting and investigating exoplanets via the transit method, which it was hugely successful at doing, discovering nearly 200 confirmed Hot Jupiter class planets between 2004 – 2018.

STING retains the basic unit-camera design of SuperWASP, but the 8-camera mosaic (which provided an instantaneous field of view of $\sim$480 deg$^2$) has been changed to 4 co-pointed units that are each equipped with a distinct fixed filter. This arrangement provides a $\sim$75 deg$^2$ field ($10.5^\circ \times 7.5^\circ$) that is observed simultaneously in four colours.

The filters currently installed in STING (shown in Figure \ref{fig:Bandpass}) are Baader CMOS R,G,B, and SDSS i. The non-standard RGB filters were selected due to their higher throughput and sharper wavelength transitions than Bessel BVR, and their better match to the sensor quantum efficiency than SDSS ugr. This is particularly important for STING, where the four cameras operate in parallel using the same exposure time and would cause the SDSS u band to have an unacceptably poor signal to noise ratio compared to the other cameras. The RGB filters are also used by the adjacent GOTO facility \citep{Steeghs2022}, making STING a complementary capability for bright astronomical targets.

\begin{figure}[H]
    \centering
    \includegraphics[width=0.5\textwidth]{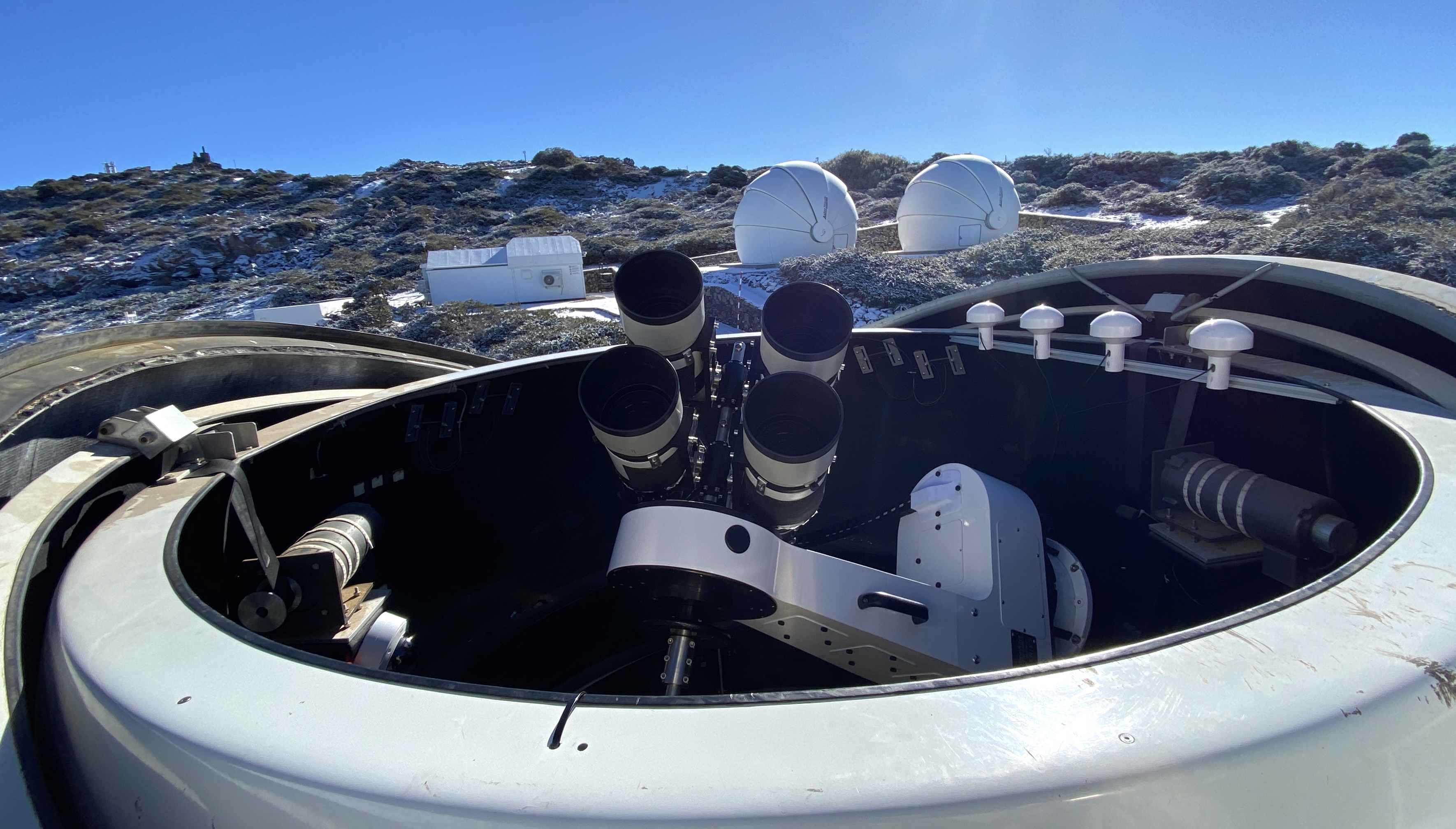}
    \caption{A photograph of the STING instrument, at the Roque de los Muchachos observatory on La Palma in the Canary Islands. The original SuperWASP enclosure and the two GOTO-North clamshell domes are visible in the background.}
    \label{fig:superwasp-photo}
\end{figure}

The CCD cameras were replaced with QHY600M-PRO CMOS cameras, which provides an improved plate scale of 3.99 arcseconds per pixel. The mount was upgraded to a direct-drive Planewave L350, and the instrument was relocated to a 7ft clamshell dome to avoid the sky visibility restrictions of the original enclosure. Both prior and new enclosures can be seen together in Figure \ref{fig:superwasp-photo}. The upgraded instrument operates robotically, controlled using a modern modular software control system (Chote, in prep).

The raw STING images are processed using an automated pipeline (described in detail by Chote et al., in prep) that applies standard image corrections (bias, dark, flat field), performs astrometric and photometric calibration, identifies targets based on an input TLE catalogue, and extracts photometric measurements with associated uncertainties. Image times are recorded through the use of GPS timestamps, and the initial exposure start time of the four cameras are synchronised in software to within a few milliseconds. The rolling electronic shutter on the CMOS sensor introduces a temporal skew of 0.25 seconds between the top and bottom rows of the image, which is accounted for in the data reduction pipeline.

Each image is individually calibrated using the star streaks visible in the image background to obtain an independent zero point and world coordinate system solution. The flux of selected reference stars (automatically determined based on their brightness and separation from nearby stars to avoid blending, typically between 100-300 per image) is matched against the  AAVSO Photometric All-Sky Survey (APASS) survey (\url{https://www.aavso.org/apass}), using a first-order color term to correct for the differences between the APASS and STING band passes. This procedure provides a robust zero point calibration that does not rely on models of atmospheric extinction.

\subsubsection{Noise Characterisation}
\begin{figure}[H]
    \centering
    \includegraphics{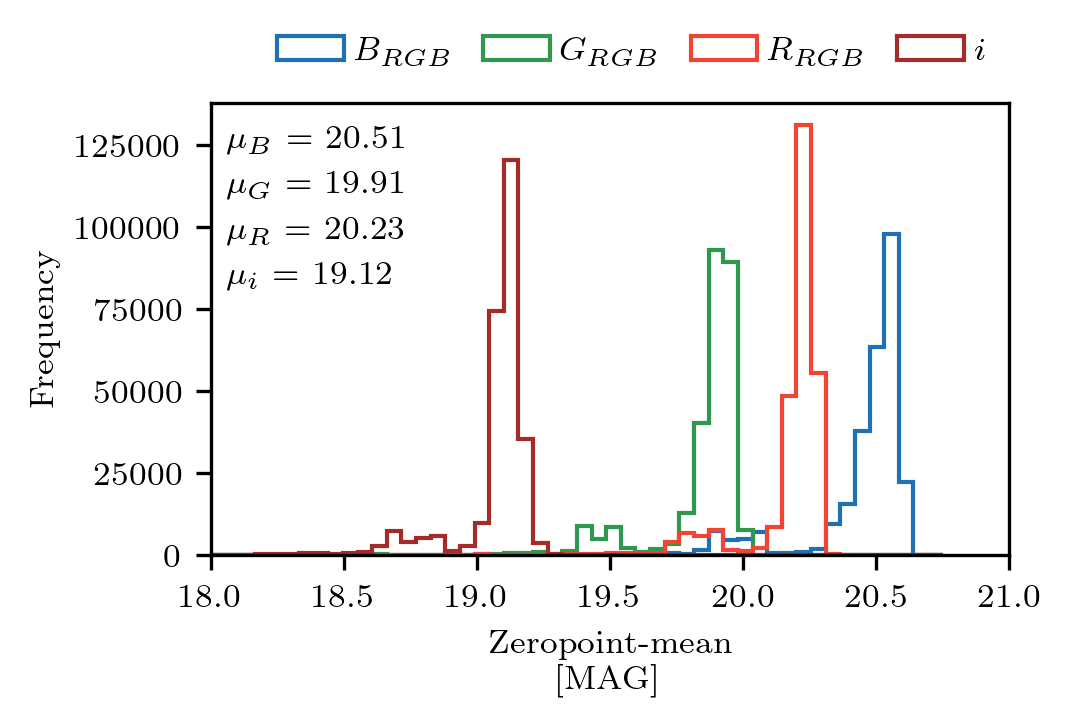}
    \caption{Distributions of the mean zeropoint values in each filter bandpass in units of magnitudes derived from flux in units of electrons. These distributions consider all the zeropoint-mean measurements across the entire observation campaign and the mean of each distribution (excluding outliers) is given in the figure legend in magnitudes. The outliers within these distributions are due to observations through thin cloud.}
    \label{fig:zp_dist}
\end{figure}
The distributions of the zeropoint-means for each of the four bandpasses are shown in Figure \ref{fig:zp_dist}.

\begin{figure*}[ht!]
    \centering
    \includegraphics[width=0.75\textwidth]{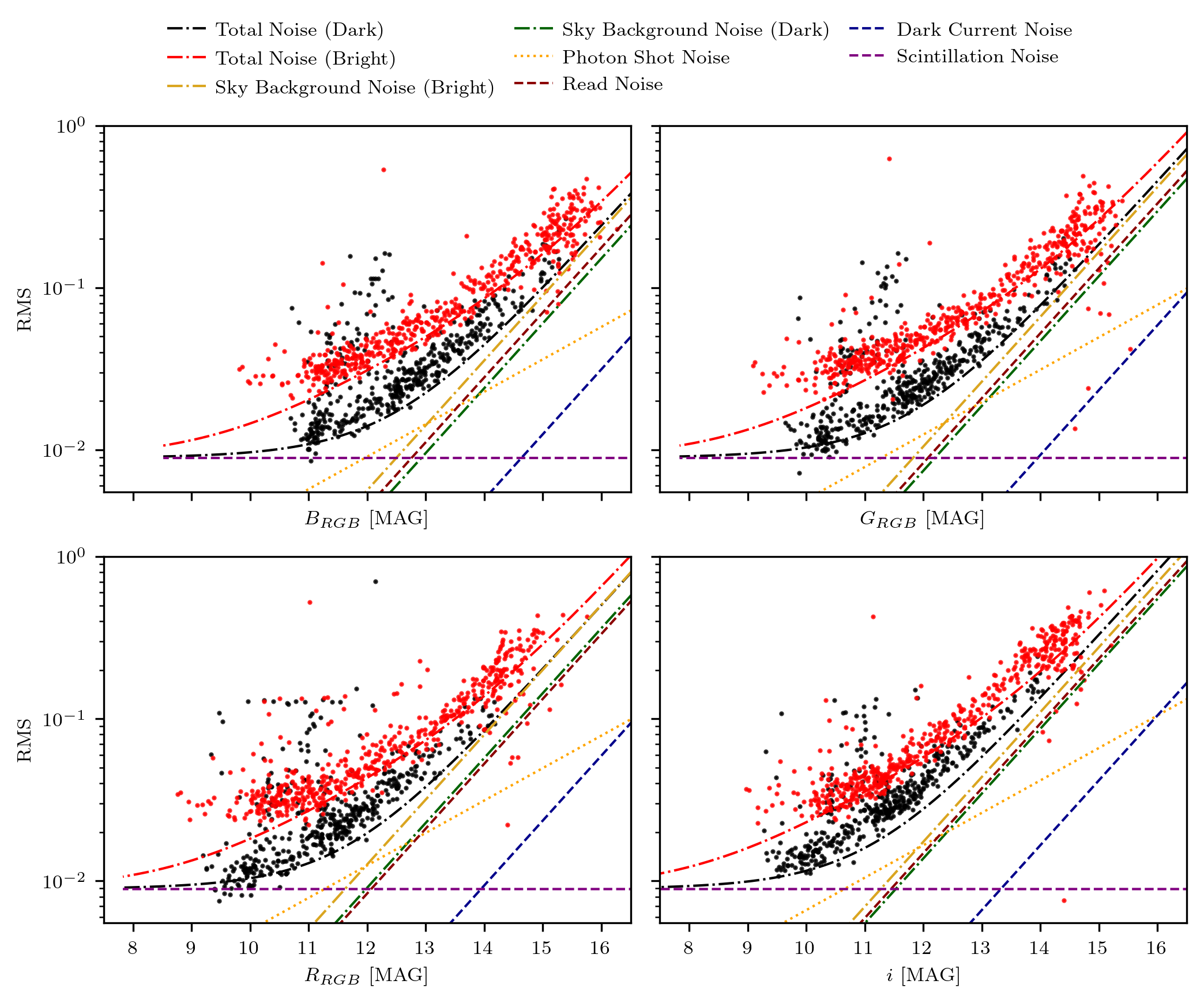}
    \caption{The measured RMS scatter in satellite light curves observed with STING agrees well with the theoretically calculated noise model in both bright and dark sky conditions. The bright time measurements are shown by the red markers with the dark time measurements being shown by the black markers. The sky background noise component was calculated on the 12th April 2023 and the 18th of May 2023 for bright and dark times respectively.}
    \label{fig:noise_characterisation}
\end{figure*}
The photometric performance of STING was characterised by constructing a theoretical noise model (see \ref{appendix:noise}) for each camera, which is shown to be consistent with the RMS scatter measured in the satellite light curves (see Figure~\ref{fig:noise_characterisation}). Satellite light curves which were measured during `bright' and `dark' nights (respective to the level of illumination of the moon disk) were binned with a sufficiently small bin width (10 minutes) such that a linear fit could be made, for which the RMS is calculated for each bin.

The typical photometric precision in a single 10\,s exposure varies between roughly 1$\%$ for targets brighter than 9th magnitude (scintillation limited), and increases to 10$\%$ at 14th magnitude (limited by sky background and readout noise). Dark current is negligible when operating the cameras at 0\,$^\circ$C. Multiple exposures can be binned to improve the precision of faint targets.

\subsection{Observation Campaign}\label{sec:obs_campaign}
Our observation campaign consisted of 17 full night observations between the 12th of April and the 18th of May 2023. Observations began at the start of nautical twilight (sun altitude -12\,$^\circ$), and continued until the end of nautical twilight the following morning. This corresponded to approximately 20:30 through 5:30 UTC during our observation campaign. Observations used a standard exposure time of 10s. 
\subsubsection{Fields and Pointing Constraints}\label{sec:Fields}
\begin{figure*}[ht!] 
    \centering
    \includegraphics[width=0.75\textwidth]{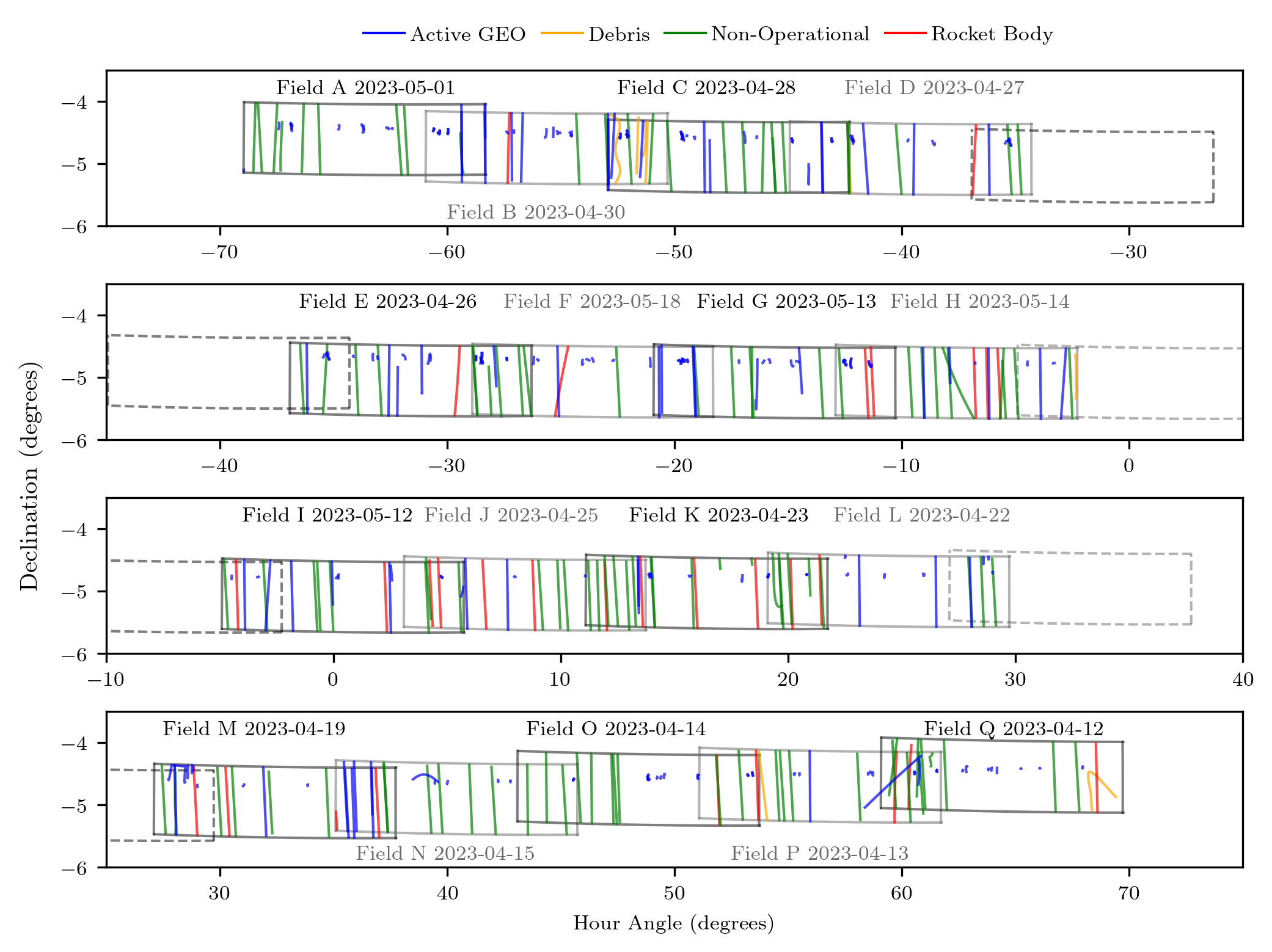}
    \caption{The observational fields (A-Q) used in this survey of the GEO belt in hour angle and declination space, with each field having overlap with the next. The tracks of each type of object we observed in our survey are displayed by their respective markers in the figure legend. The borders of each field are alternated between black and grey in order to distinguish between them. Dashed field lines are drawn to display the fields on the previous/next row}
    \label{fig:An example_FIELD_TILED}
\end{figure*}
The GEO belt was tiled into 17 fields spanning hour angles between $\pm$\,70\,$^\circ$. A $\sim2.5\,^\circ$ overlap between adjacent fields ensures that the GEO targets of interest remain within the field of view of all four cameras for the entire night of observing (accounting for the small misalignment in the camera footprints, and the motion of targets throughout an observation series). The vertical field of view was cropped to 1\,$^\circ$ to reduce the data volume; this is sufficient for observing active objects that maintain low-inclination orbits, but sacrifices observation duration for debris and objects in inclined orbits (measured to have a duration at most around 170 minutes) as they pass through the limited vertical field. 
A single field was observed robotically each night, chosen to ensure that the moon did not approach within 30\,$^\circ$ of the field centre, and where possible to avoid the galactic plane (with very high background star density) entering the field during the observation. Due to visibility constraints from the walls of the enclosure, some fields were unable to be observed with some cameras, such was the case for the $R_{RGB}$ camera observations in Field B. In this case, we neglected the observations taken by the camera. The observational fields and objects observed within are displayed in Figure \ref{fig:An example_FIELD_TILED}.

\section{Results}\label{sec:Results}

\begin{figure*}[ht!]
    \centering
    \includegraphics{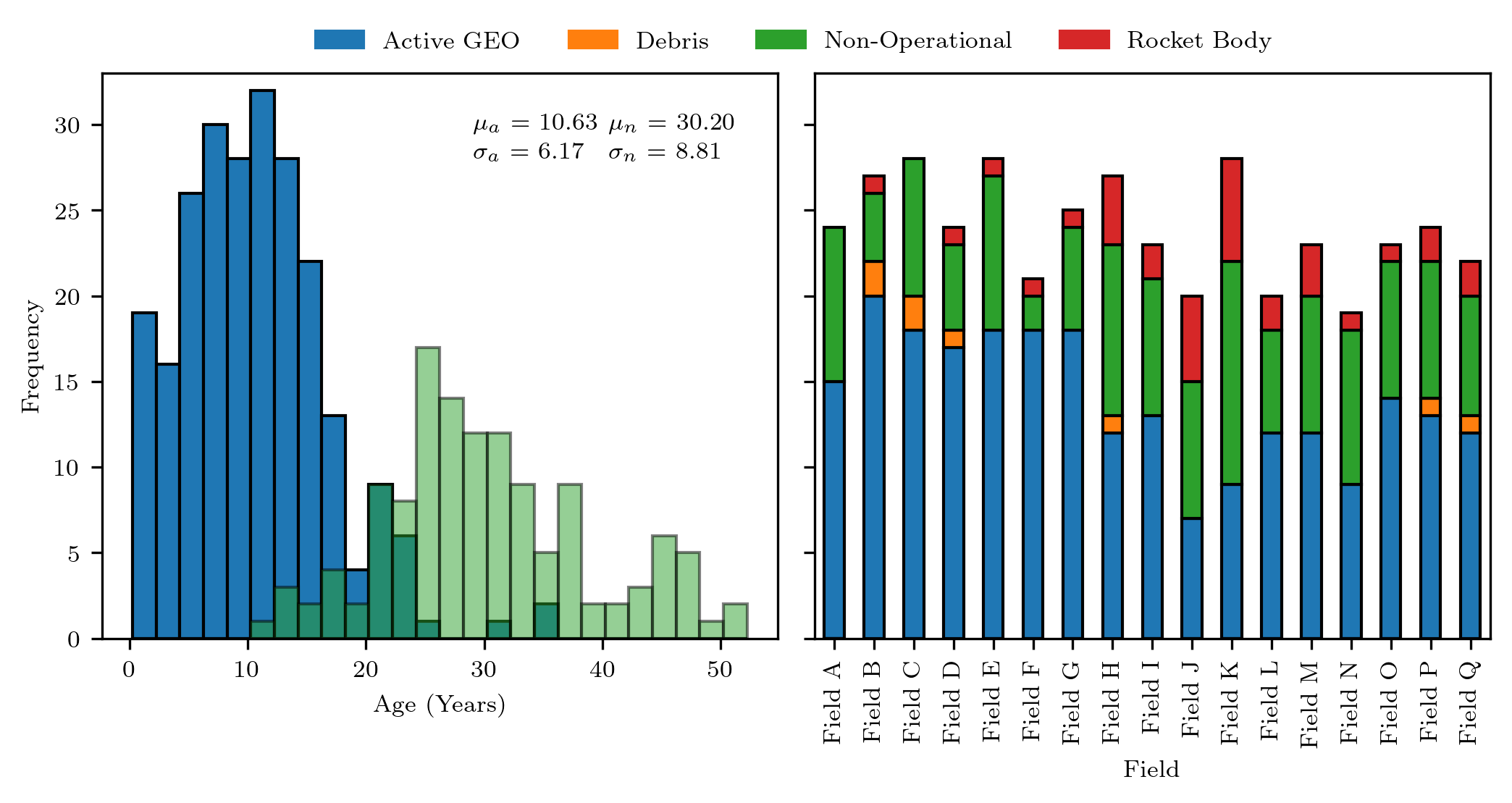}
    \caption{Demographics of observed objects upon conclusion of our survey. Left panel: Histogram displaying the overlapping age distributions of observed active and non-operational geostationary satellites in 2 year bins. Here $\mu_a$, $\sigma_a$ and $\mu_n$, $\sigma_n$ refer to the mean and standard deviation of the active and non-operational satellites respectively. A transparency is applied to the distribution of non-operational satellites. Right panel: Stacked bars indicating the type and number of objects observed within each field of our survey.}
    \label{fig:DEMO1}
\end{figure*}

\subsection{Survey Demographics}

The majority of geostationary satellites feature a `box-wing' design, with large flat solar panels that have freedom to pivot around the north-south axis in order to track the sun. Satellites of this type typically display strong specular features around solar equatorial phase angle of 0\,$^\circ$ due to contributions of specular reflection from the solar panels \citep{Payne2007}.

Our observations were carried out in the weeks following the March equinox, where the Sun-satellite-observer plane aligns with the rotation axis of the Earth. This means that the component of the phase angle perpendicular to the equator is also close to zero, leading to greater specular reflections from the solar panels -- the so called `glint season' \citep{2016amos.confE.126W}. The geometry following the `glint season' remains favourable for detecting specular reflections from the satellites without suffering from the extreme brightness changes that would otherwise saturate the detectors during the brightest glints.

The bus of a GEO satellite will typically have a fixed orientation relative to the Earth so as to maintain a consistent antenna footprint. Other light curve features can be possibly explained by reflections of the satellite bus or mounted components on said bus. 
\begin{table}[H]
    \centering
    \begin{tabular}{c|cccc|c}
    %\hline
    Field & & &  &\\
    ID &  ACT & DEF & RB & DEB & Total\\ \hline
    A &  15 & 9 & 0 & 0 & 24 \\ 
    B &  20 & 4 & 1 & 2 & 27 \\ 
    C &  17 & 9 & 0 & 2 & 28 \\ 
    D &  17 & 5 & 1 & 1 & 24 \\ 
    E & 18 & 10 & 1 & 0 & 29 \\
    F &  19 & 2 & 1 & 0 & 22 \\ 
    G & 17 & 7 & 1 & 0 & 25 \\
    H &  11 & 11 & 4 & 1 & 27 \\ 
    I &  11 & 9 & 2 & 0 & 22 \\ 
    J & 7 & 8 & 5 & 0 & 20 \\ 
    K &  9 & 13 & 6 & 0 & 28 \\ 
    L &  11 & 7 & 2 & 0 & 20 \\ 
    M &  12 & 8 & 3 & 0 & 23 \\ 
    N &  9 & 9 & 1 & 0 & 19 \\ 
    O &  14 & 8 & 1 & 0 & 23 \\ 
    P &  13 & 8 & 2 & 1 & 24 \\ 
    Q &  12 & 7 & 2 & 1 & 22 \\ \hline
    Total Observations  & 232 & 134 & 33 & 8 & 407 \\\hline
    Unique Objects & 168  & 109 & 22 & 6  & 305\\
    \end{tabular}
    \caption{Summary of objects observed in each of the 17 survey fields. The different types of object can be broken down into active geostationary satellites (ACT), non-operational satellites (DEF), rocket bodies (RB) and debris (DEB). Objects may be observed in adjacent fields if it lies within the overlapping region, this means that the unique total of objects is less than the sum of observations. }
    \label{tab:Object Type Count}
\end{table}
\vspace{1cm}
Our analysis focuses on the active GEO satellite population, which spans a wide distribution of time on orbit. This allows the possible relationship between time on orbit and reddening of satellites over time due to material degradation to be investigated. Figure \ref{fig:DEMO1} and Table \ref{tab:Object Type Count} provide an overview of the full observational sample. We observed 305 unique objects in total, 168 of these were active satellites. For objects at lower elevations (fields A, B, and Q), one or more of the cameras were obscured by the dome horizon and those specific colour bands are entirely absent. The 15 active satellites observed in Field A were missing two colour bands so were excluded. 37 active satellites on inclined orbits were excluded due to insufficient observation length and 4 satellites which were missing at least one colour band were excluded also, such that we are left with our final sample of 112 objects. Table \ref{tab:Bus Configuration Count} displays the different types of bus configuration associated with the final 112 objects.

\begin{figure*}[ht!]
    \centering
    \includegraphics{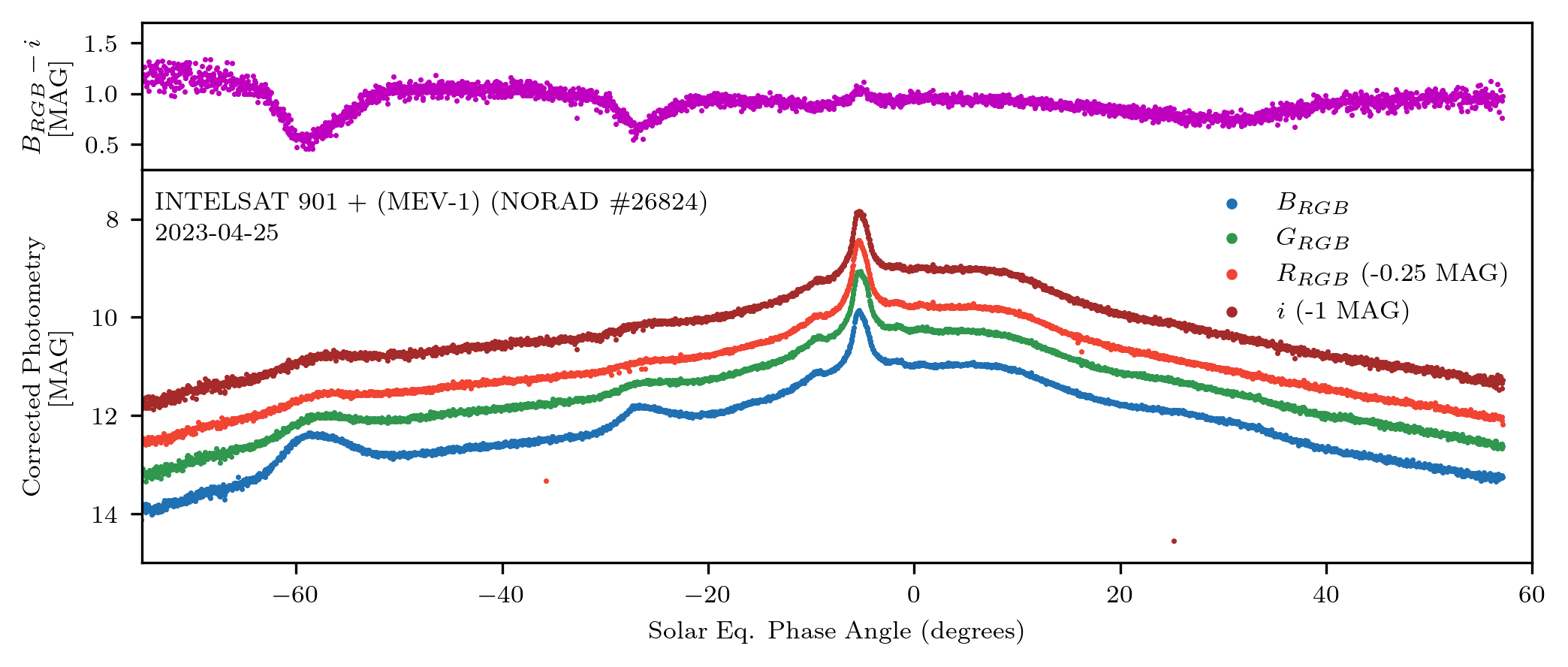}
    \caption{An example four colour light curve of INTELSAT 901 (+ MEV-1) on the 25th April 2023. The upper panel displays the $B_{RGB} - i$ colour index with the lower panel displaying the corrected photometric measurements in the four colour bands. A 0.25 and 1 mag offset is applied to the $RGB_{R}$ and $i$ band measurements respectively to aid readability.}
    \label{fig:lc_eg}
\end{figure*}
\begin{table}[H]
    \centering\footnotesize
    \begin{tabular}{rl}
        Bus Configuration & Count \\ \hline
        Eurostar-3000 & 26 \\
        SSL-1300 & 24 \\
        STAR-2 & 10 \\
        BSS-702HP & 5 \\ 
        Spacebus-4000B2 & 4 \\
        BSS-702MP & 4 \\ 
        Spacebus-4000C3 & 3 \\ 
        Spacebus-Neo-200 & 3 \\
        Spacebus-4000C4 & 3 \\
        A2100AXS & 3 \\ 
        GEOSTAR-3 & 2 \\
        Eurostar-NEO & 2 \\ 
        Spacebus-4000B3 & 2 \\
        LUXOR & 2 \\ 
        Ekspress-2000 & 2 \\ 
        Eurostar-2000+ & 2 \\ 
        Spacebus-3000B3 & 2 \\ 
        Spacebus-3000B2 & 1  \\ 
        Spacebus-Neo-100 & 1 \\ 
        LM-2100 & 1 \\ 
        SS-702 & 1 \\
        AMOS & 1 \\ 
        DS2000 & 1 \\
        Ekspress-1000N & 1 \\ 
        Eagle-Bus ESPA (ESPAStar-D) & 1 \\
        BSS-702SP & 1 \\
        Ekspress-1000HTB & 1 \\ 
        ARSAT-3K & 1 \\ 
        DFH-4 & 1 \\ 
        A2100A & 1 \\ \hline
        Grand Total & 112
    \end{tabular}
    \caption{Count of bus configurations for the unique active GEO sample of satellites for which photometry was available in all four bands ($B_{RGB}$, $G_{RGB}$, $R_{RGB}$ and $i_{sdss}$).}
    \label{tab:Bus Configuration Count}
\end{table}
\begin{figure*}[p!]
    \centering 
    % First figure
    \includegraphics[width=\textwidth]{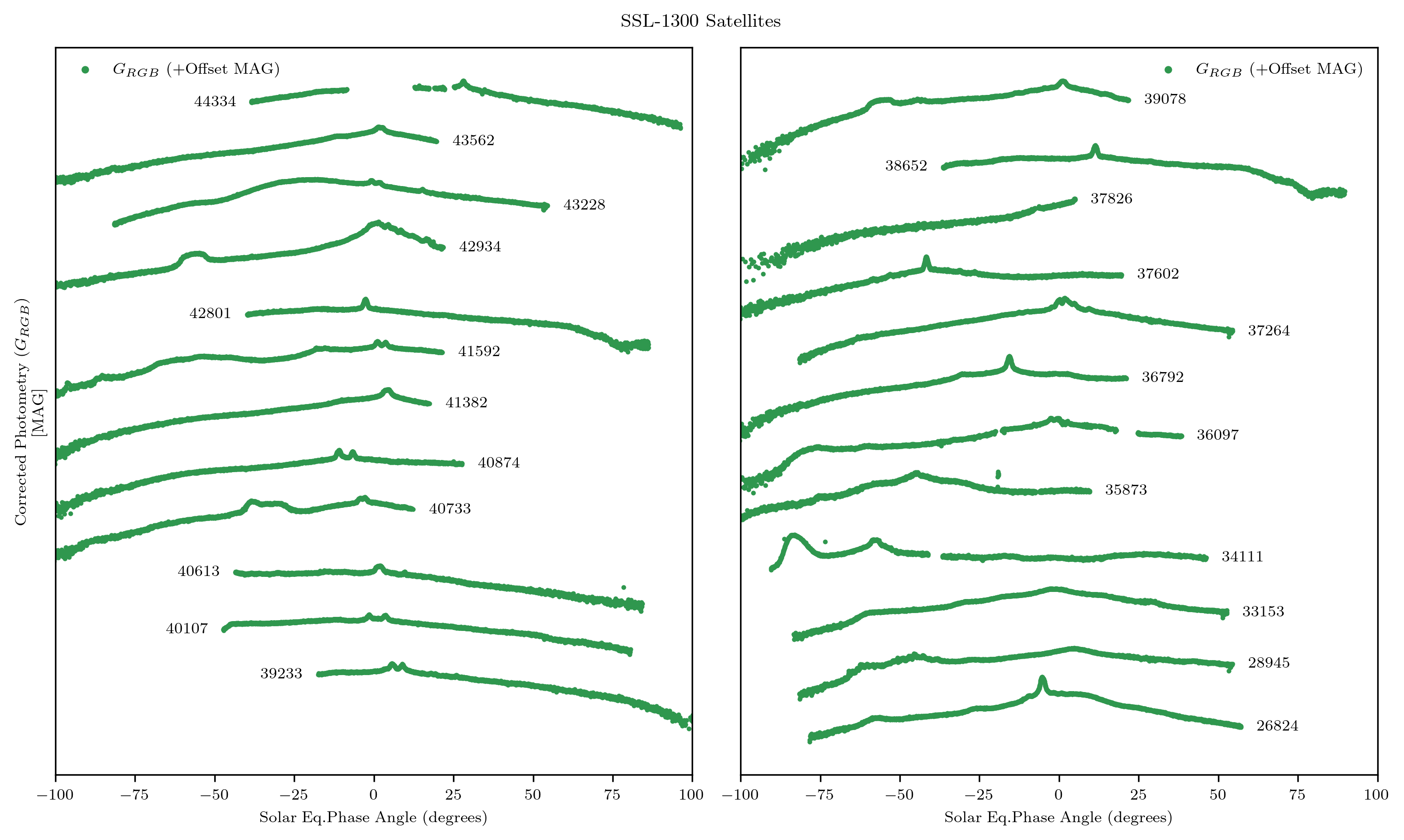}
    \caption{Illustrative example of light curves for SSL-1300 class satellites with an offset of 4 magnitudes applied to each subsequent light curve to avoid overlap. Gaps within a few of the light curves are due to filtering out bad zeropoint measurements. The NORAD ID of each object is shown adjacent to its light curve.}
    \label{fig:SSL_eg}
    
    \vspace{0.5em} % Add some vertical space between the figures
    
    % Second figure
    \includegraphics[width=\textwidth]{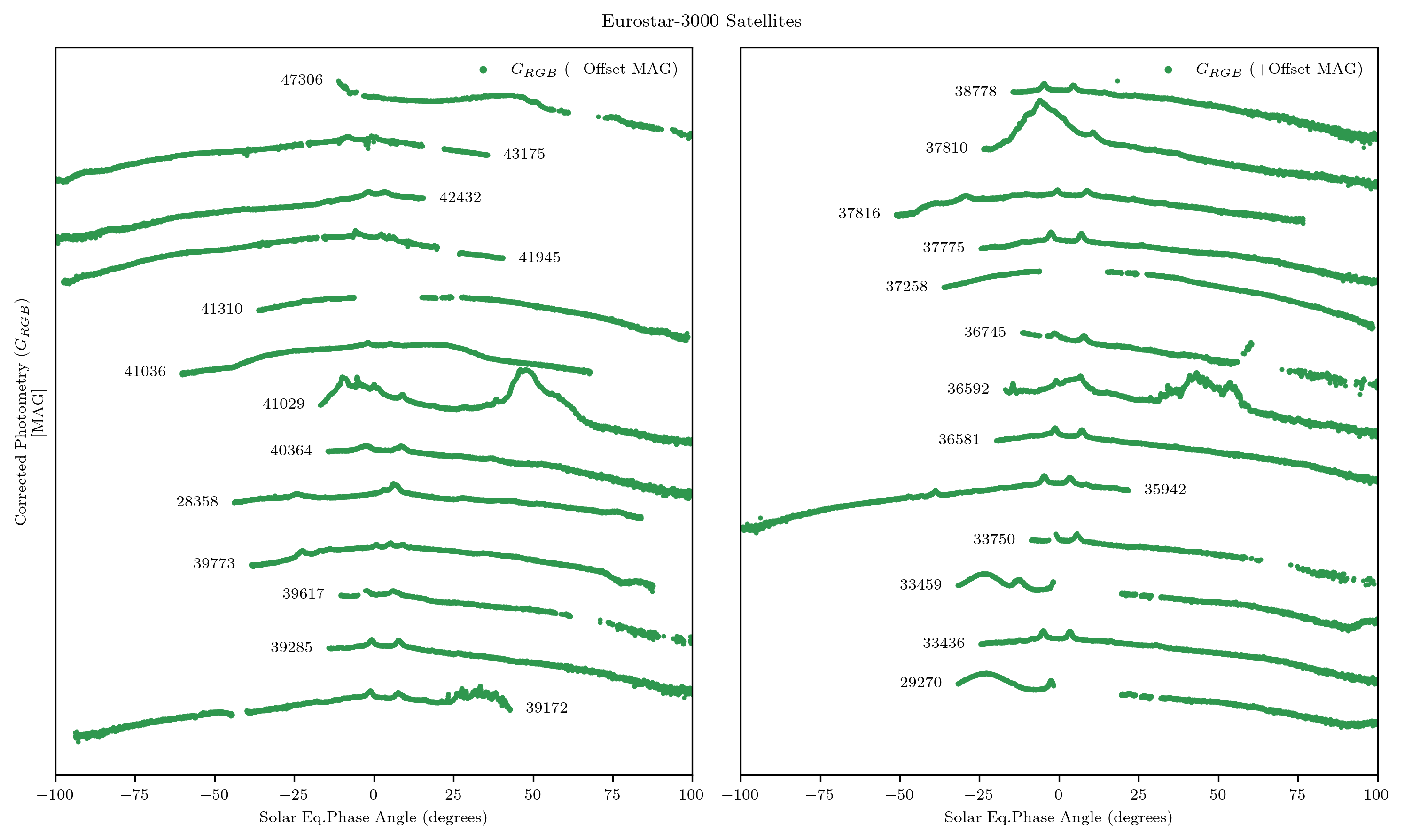}
    \caption{Illustrative example of light curves for Eurostar-3000 class satellites with an offset of 4 magnitudes applied to each subsequent light curve to avoid overlap. Gaps within a few of the light curves are due to filtering out bad zeropoint measurements. The NORAD ID of each object is shown adjacent to its light curve.}
    \label{fig:e3000_eg}
    
\end{figure*}
\subsection{Phase Angle Light Curves}
The profile and features within our light curves are the result of sunlight reflected from the satellite's surface in the direction of the observer. The most useful way to capture this geometry is through the concept of the solar equatorial phase angle \citep{Payne2007}. The solar equatorial phase angle is defined as the angle between the projection of sun and observer on the celestial equator \citep{XiaoFen2021LongtermPS} or moreover corresponds to the the difference in Right Ascension between the satellite and anti-solar point (the point directly opposite the Sun on the celestial sphere from the observer perspective).

Figure \ref{fig:lc_eg} displays an example of a processed four colour light curve of the INTELSAT 901 (+ MEV-1) satellite observed on the 25th April 2023 with STING. This observation is taken after the docking of INTELSAT-901 with MEV-1 and therefore the colour signatures within could potentially be used to distinguish between INTELSAT-901 and MEV-1 components. There are notable colour features within this light curve which will relate to objects that we analyse in later sections, particularly section \ref{sec:7CC}. This data demonstrates the ability of the STING instrument to resolve these signatures.
 
Figures \ref{fig:SSL_eg} and \ref{fig:e3000_eg} present a selection of $G_{RGB}$ band light curves of satellites belonging to the SSL-1300 and Eurostar-3000 bus classes respectively.

\section{Analysis}\label{sec:Analysis}
\subsection{Induced Colour Error}
Sequential colour measurements can induce an error when the object brightness is rapidly changing with respect to the filter cycle, this is referred to an induced colour index error \citep{castro2022Analysis}.

We wish to demonstrate the improvements in the resolution of light curve features achieved through simultaneous colour measurements, as compared to the sequential colour measurements predominantly used in previous studies. We analyse how and where simultaneous observations enhance the resolution of features within the light curves.

\begin{figure}[H]
    \centering
    \includegraphics{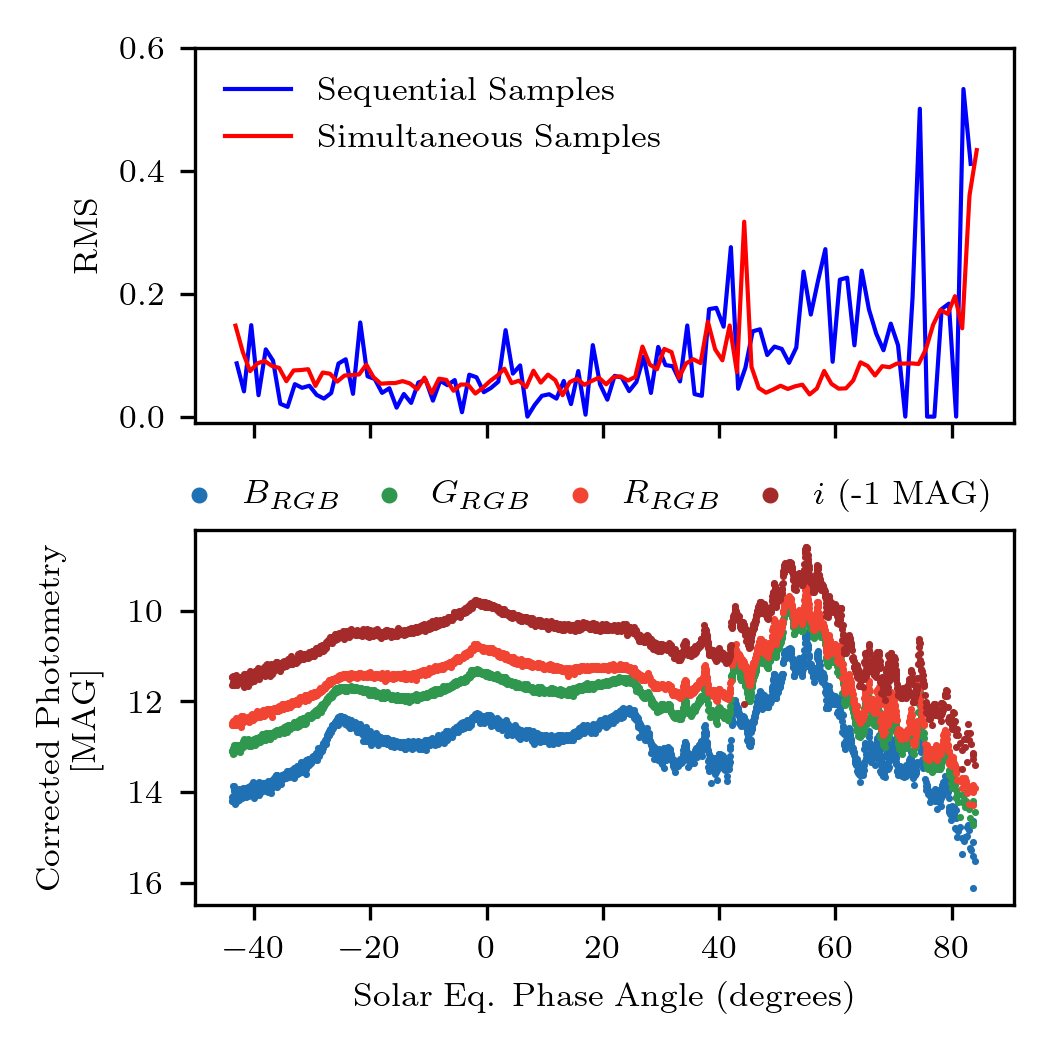}
    \caption{Upper: Binned RMS values calculated for the sequential and simultaneous measurements of $B_{RGB} - i$ within bins of width of 5 minutes. Lower: Original THOR-6 light curve with truth simultaneous measurements. The sequential measurements have a definitively larger RMS over the glinting region due to not being able to effectively resolve these short timescale features. A 1 mag offset is applied to the $i$ band measurements.}
    \label{fig:thor-6-seq-sim}
\end{figure}

Figure \ref{fig:thor-6-seq-sim} is an example of the light curve of THOR-6 with a binned RMS measurement taken in bins with widths of 5 minutes. For the majority of the light curve, the two colour features have a negligible deviation difference when comparing the simultaneous and sequential observations. However, as we approach the region between 40-75\,$^\circ$, where the brightness changes on short timescales, we see there is a clear larger associated RMS with the sequential data measurements, this highlights the importance of simultaneity such that simultaneous colour measurements are most useful across short timescale features and these features in particular are likely to have interesting colour behaviours that we wish to investigate.
\begin{table}[H]
    \centering
    \begin{tabular}{cccc}
        NORAD & $\sigma_{seq}$ & $\sigma_{sim}$ & $\Delta$ \\
        ID & (M) & (M) & ($\%$)\\ \hline
         42934 & 0.113 & 0.086 & 27.1 \\ 
         29273 & 0.103 & 0.093 & 10.6 \\ 
         41866 & 0.105 & 0.112 & -6.0 \\ 
         39168 & 0.082 & 0.064 & 24.4 \\ 
         39172 & 0.211 & 0.103 & 68.5 \\
         28945 & 0.156 & 0.105 & 39.0 \\
         45027 & 0.123 & 0.124 & -0.50 \\ 
        44457 & 0.161 & 0.102 & 44.2 \\ 
        37836 & 0.222 & 0.188 & 16.7 \\ 
        41029 & 0.255 & 0.188 & 30.4 \\ 
        36592 & 0.212 & 0.145 & 37.5 \\ 
         40258 & 0.136 & 0.072 & 61.9 \\ 
         36830 & 0.183 & 0.149 & 20.9 \\ 
        36033 & 0.307 & 0.229 & 29.2 \\ 
         36831 & 0.13 & 0.098 & 27.5 \\ 
         39163 & 0.443 & 0.367 & 18.7 \\\hline
    \end{tabular}
    \caption{Standard deviation measurements for both sequential and simultaneous colour measurements ($B_{RGB} - i$) of short timescale glinting regions within the light curves of the named satellites. The standard deviation measurements are in units of magnitudes (M). The percentage difference between the standard deviations is denoted as $\Delta$. The percentage difference is calculated as $\frac{a-b}{0.5(a+b)}$, where $a$ and $b$ are the standard deviation measurements of the sequential and simultaneous samples respectively. }
    \label{tab:ci_error_glints}
\end{table}

Given we had numerous objects in our dataset which show regions of features occurring on short timescales, we characterised and measured the induced colour index error associated with the sequential data measurements. Table \ref{tab:ci_error_glints} shows the measured standard deviation of the colour index ($B_{RGB} - i$) for various satellite glinting regions for both simultaneous and sequential observations. Generally, we find that the standard deviation increases significantly between the simultaneous and sequential data measurements. The average percentage difference and increase between the simultaneous and sequential measurements is approximately 28$\%$ and 36$\%$ respectively. The GOES 16 (NORAD: 41866) and EUTE KONNECT (NORAD: 45027) satellites actually show a reduction in the error, however, this can be explained as the glinting seen within their light curves are occurring over slightly longer timescales than the others (see Section \ref{sec:glinting}). It is evident that the use of simultaneous colour measurements allow for a higher resolution of these glinting regions.

\subsection{Colour Contrasts Between Bus Configurations}

\begin{figure*}[p!]
    \centering
    \includegraphics[width = \textwidth,height = \textheight]{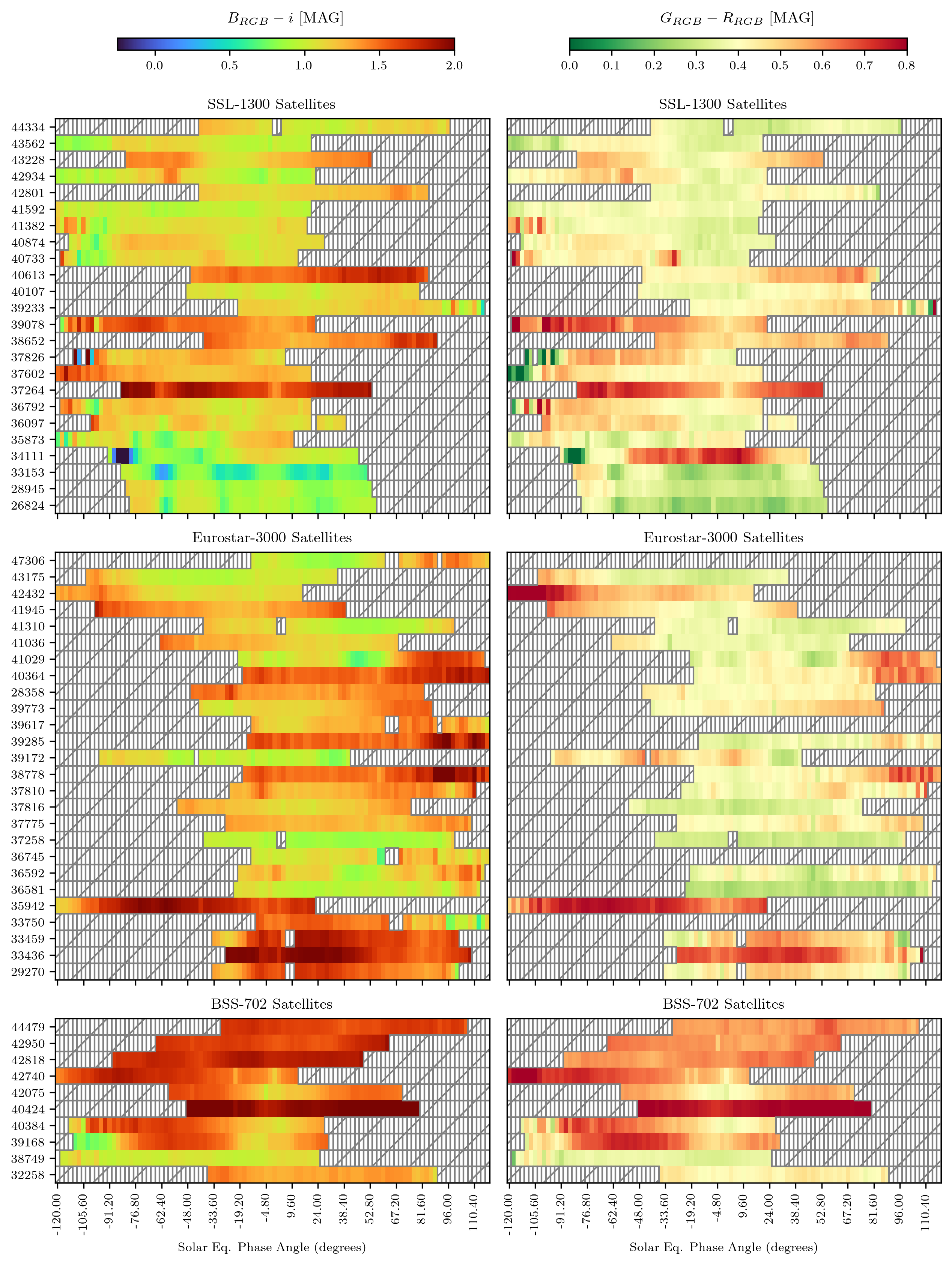}
    \caption{Colour light curves transformed into 2D image data for $B_{RGB} - i$ and $G_{RGB} - R_{RGB}$ colour indices. Each row represents the colour light curve of a satellite object identified by its given NORAD id and each block of objects belong to a particular  bus configuration. Each colour light curve is split into 100 bins of equal solar equatorial phase angle width. The row order is determined by increasing time on orbit and entire missing rows correspond to objects observed in field B.}
    \label{fig:bus_ims_1}
\end{figure*}

\begin{figure*}[p!]
    \centering
    \includegraphics[width = \textwidth,height = \textheight]{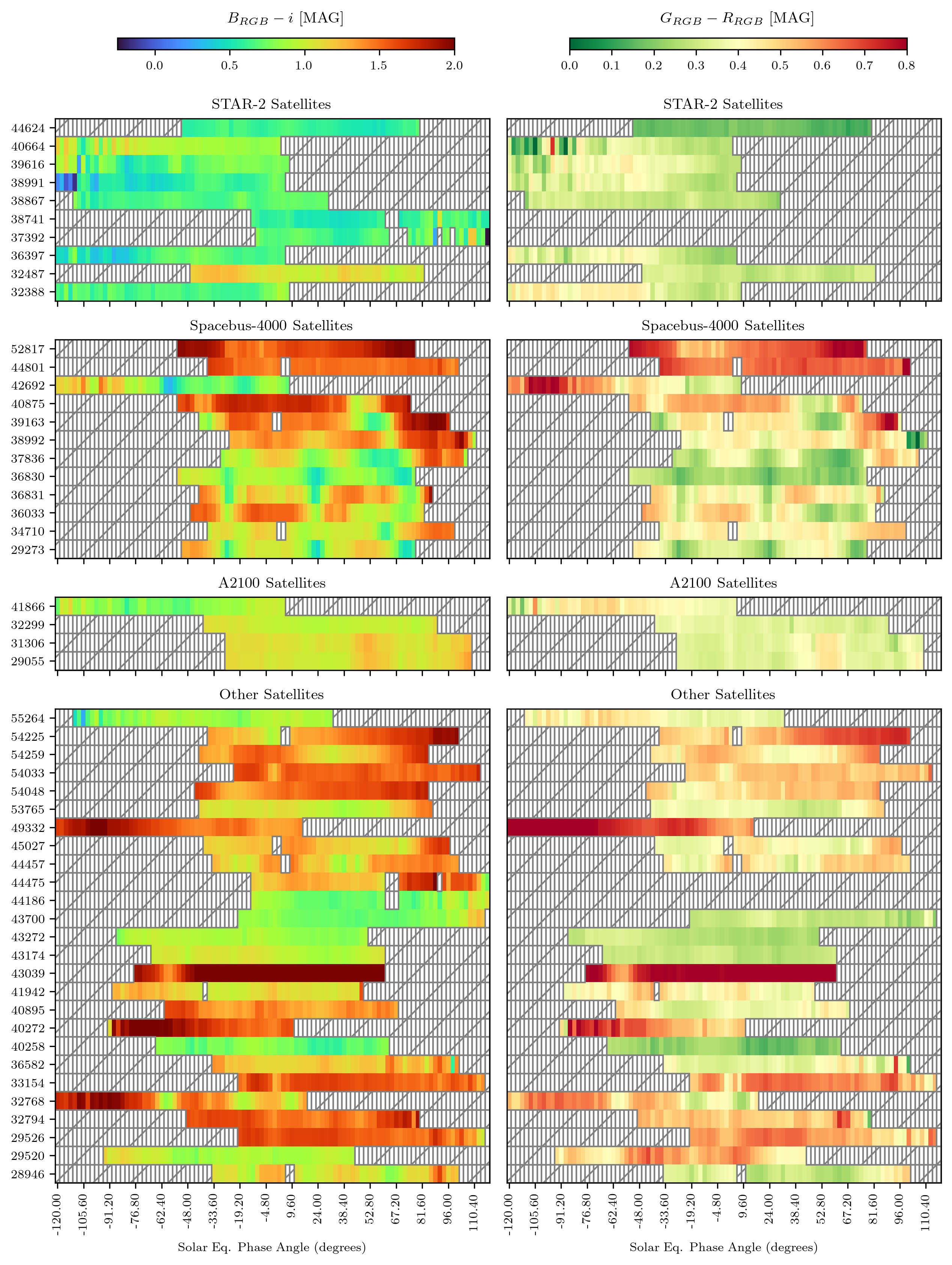}
    \caption{Colour light curves transformed into 2D image data for $B_{RGB} - i$ and $G_{RGB} - R_{RGB}$ colour indices. Each row represents the colour light curve of a satellite object identified by its given NORAD id and each block of objects belong to a particular  bus configuration. Each colour light curve is split into 100 bins of equal solar equatorial phase angle width. The row order is determined by increasing time on orbit and entire missing rows correspond to objects observed in field B.}
    \label{fig:bus_ims_2}
\end{figure*}
To further analyse the observed differences between satellites of different bus configurations, we took the approach of investigating two colour indexes ($B_{RGB} - i $ and $G_{RGB} - R_{RGB}$) as a function of solar equatorial phase angle for each satellite. A consideration for this again is to exclude the $G_{RGB} - R_{RGB}$ measurements for satellites observed within field B (large variability in the $R_{RGB}$ zeropoint).

We chose the phase range of -120\,$^\circ$ to 120\,$^\circ$ as this captured most if not all of the light curve datapoints of our sample targets. We then split our colour light curves into 100 bins ($\simeq$ 10 minute bin width, small enough to not lose any resolution on significant features). We could then map these to a suitable colour map range for $B_{RGB} - i $ and $G_{RGB} - R_{RGB}$ respectively, ensuring comparison between every bus configuration and produce the 2D image plots shown in Figure \ref{fig:bus_ims_1} and Figure \ref{fig:bus_ims_2}. These colour light curve maps are particularly useful for displaying the variability of colour across an observation as well as making the identification of features shared between objects easier.

The colours of GEO satellites are expected to vary both across different bus configurations and throughout the night for individual objects. Starting with the SSL-1300 class satellites, we can comment on a few interesting features. Generally the value of $B_{RGB} - i$ tends to be fairly consistent for individual satellites that were launched more recently and this value of colour index is comparable to that seen in Eurostar-3000 and A2100 satellites. For the objects which have spent more time on orbit, there tends to be `bluer' features present and in general more variability in the $B_{RGB} - i$ index values. The value of the $G_{RGB} - R_{RGB}$ colour index shows greater variability across all SSL-1300 satellites, however the same cluster of objects with the intense $B_{RGB} - i$ excursions (excursion referring to an increase or a decrease in the colour index value) are `greener' than the other SSL-1300 satellites.

The Eurostar-3000 satellites display an effect of the satellites which have spent the most time on orbit displaying much `redder' colours with respect to satellites of the same bus class that were launched more recently. The Eurostar-3000 satellites launched after AMAZONAS-2 (NORAD: 35942) don't reach the same values of $B_{RGB} - i$ consistently. However, looking at the $G_{RGB} - R_{RGB}$ index we see that whilst Eurostar-3000 satellites are `greener' with respect to the cluster of objects launched before ASTRA-3B (NORAD: 36581), the amount of colour difference between this group and the others is not as drastic as shown in the $B_{RGB} - i$ index. 

BSS-702 satellites tend to be `redder' across the entire phase range. We also see this behaviour in the $G_{RGB} - R_{RGB}$ colour index. Contextually, their $B_{RGB} - i$ indices are comparable to that of the cluster of red Eurostar-3000 satellites. Conversely, STAR-2 satellites are consistently `bluer' across the entire phase range than any of the other satellite platforms within the $B_{RGB} - i$ indices. The $G_{RGB} - R_{RGB}$ values tend to be only marginally smaller than satellites within other bus classes. 

In the Spacebus-4000 class, a few satellites tend to show similar colour features at similar solar equatorial phase angles with some near-symmetry between features around $\pm$ 25\,$^\circ$. These features are apparent in both colour indices and are instances in which the satellite appears `bluer'. Conversely, the sample size of A2100 satellites is small however we can see that generally the colour of these satellites tends to be fairly consistent across the entire phase range. The two older satellites (NORAD IDs: 31306 and 29055) show a red feature around 60\,$^\circ$ of phase. The A2100 also display slightly `greener' colours compared to other satellite bus configurations.
\begin{figure*}[ht!]
    \centering
    \includegraphics{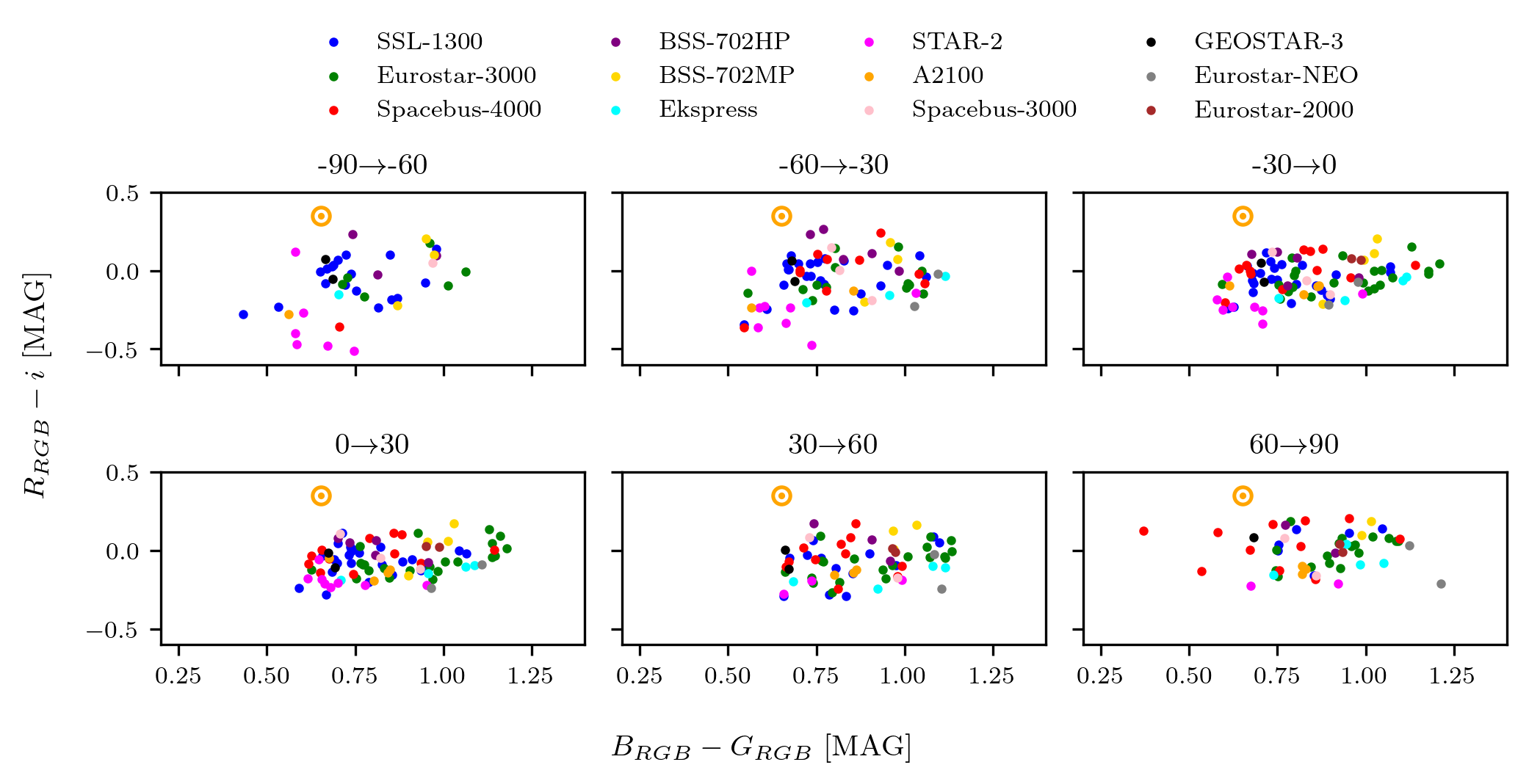}
    \caption{Colour-colour diagrams showing the average colour measurements of individual satellites belonging to a particular bus configuration class within given solar equatorial phase angle ranges from  $\pm$ 90\,$^\circ$. The error-bars were significantly small on these measurements, so they are not displayed. The average colours of the sun are denoted by the orange solar marker on each of the diagrams.}
    \label{fig:Col_COL}
\end{figure*}

\subsection{Colour-Colour diagram}
Figure \ref{fig:Col_COL} displays the average colour measurements of satellites belonging to a specific bus class within a specific phase angle range of their light curves. There is significant overlap in the colour measurements for many of the bus configurations, which indicates characterisation maybe difficult based on the colour alone for identification of many of the bus classes. In particular, as the phase angle varies between $\pm$\,30\,$^\circ$, the colours of the satellites of varying bus configurations cluster closely together. This highlights the difficulty in using colour measurements to discriminate between bus configurations overall. 
The colour indices ($B_{RGB} - G_{RGB}$ and $R_{RGB} - i$) were chosen as they best demonstrate the behaviour seen between classes at the very blue ($\sim400-550$ nm) and very red ($\sim 600-850$ nm) regimes across each axis respectively.

We also note that all of the satellite colour measurements overall have `redder' colours than the solar spectrum, which is in agreement with the measurements from \cite{SCHMITT2020326}.

\subsection{Light Curve Features}\label{sec:7CC}
\begin{figure*}[h!]
    \centering    \includegraphics{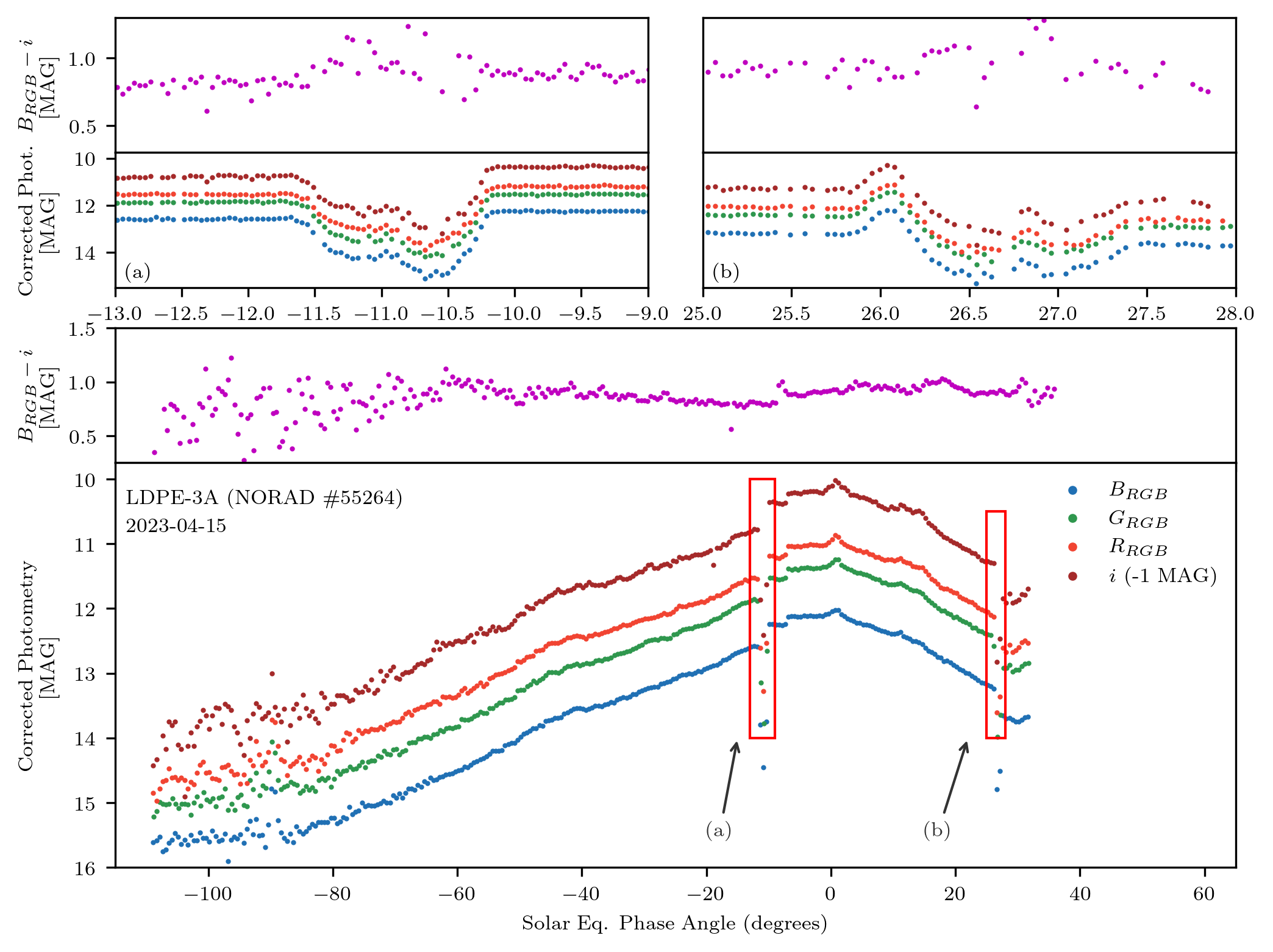}
    \caption{Binned (bin widths of 7.5 minutes) four colour phase light curves for satellite LDPE-3A (NORAD: 55264) observed on the 15th April 2023. At this point, LDPE-3A had only spent 90 days on orbit. Interesting features are labelled by \textbf{(a)} and \textbf{(b)}. We suspect that each of these features are caused by the manoeuvring of LDPE-3A about its own axes. The upper two panels show the zoomed in and unbinned region of the light curve for features annotated by \textbf{(a)} and \textbf{(b)} respectively with these regions being shown by the boxes drawn in the lower panel. We also note slight stepped features across the phase light curve which further suggest slight manoeuvring. A 1 mag offset is applied to the $i$ band measurements for clarity.}
    \label{fig:ldpe3a}
\end{figure*}
\subsubsection{LDPE-3A Manoeuvrers}
LDPE-3A (ROOSTER-3) is a satellite designed with two experimental payloads for space-domain awareness \citep{Gunther-SpacePage}. The first is named Catcher, which is designed to demonstrate new miniaturised technology capable of diagnosing the adverse effects of radiation, charged particles, and other space weather events on spacecraft in orbit. The second is WASSAT, this is a wide-area sensor with four cameras that searches for and tracks other spacecraft and space debris in geosynchronous orbit, where communications, missile detection, intelligence-gathering and weather monitoring satellites operate. It also has the most unique design of all satellites within our dataset given it is a ring-shaped spacecraft with a singular solar panel, LDPE-3A and GOES 16 are the only satellites in our sample with singular solar panel designs.

The phase and colour light curves of LDPE-3A are presented in Figure \ref{fig:ldpe3a} and denoted are several interesting features. Features denoted by \textbf{(a)} and \textbf{(b)} are likely body manoeuvres of LDPE-3A in such a way that the area of LDPE-3A presented to STING has decreased such that we see a sharp decrease in brightness. Although the differing factors between both features is that we see a slight increase in the brightness before the sharp decrease in \textbf{(b)}, which is shown by the zoomed inset axis placed in Figure \ref{fig:ldpe3a}.

We find that both of these features are associated with slight shifts towards `redder' colours which could be reflections off the gold MLI covering of some of the instruments. Although \textbf{(b)} does show a shift towards `bluer' colours whilst the brightness decreased. The inset axes which holds the unbinned datapoints of this portion of the light curve does reveal that during the decrease in brightness, there is a short lived glint where the brightness increases which suggests subtle changes in the attitude, i.e. more surface area is in the line of sight of STING. Following \textbf{(a)} we see that there is a slight step up in the light curve which suggests again a slight attitude adjustment of LDPE-3A. It would also suggest that this isn't a station-keeping manoeuvre such that rather than returning to the same orientation it now lies in a different orientation. Given that the LDPE series of satellites have instruments aboard for space domain awareness, it could be possible that it is now tracking a different object.

Overall there isn't much variability in the colour and the colour could be comparable to that of the STAR-2 satellites. Given that these were designed by the same manufacturer (Orbital Science Corp, later Northrop Grumman), it is possible that the same blanket materials were used.

Similar signatures to \textbf{(b)} can also be seen in the light curves of the MEV-2 satellites during proximity operations with Intelsat 10-02 (Chote et al., in prep). The features identified in this light curve could be useful for future identifications of manoeuvres in other light curves of GEO satellites.

\subsubsection{Blue Colour Features}
\begin{figure*}[ht!]
    \centering
    \includegraphics{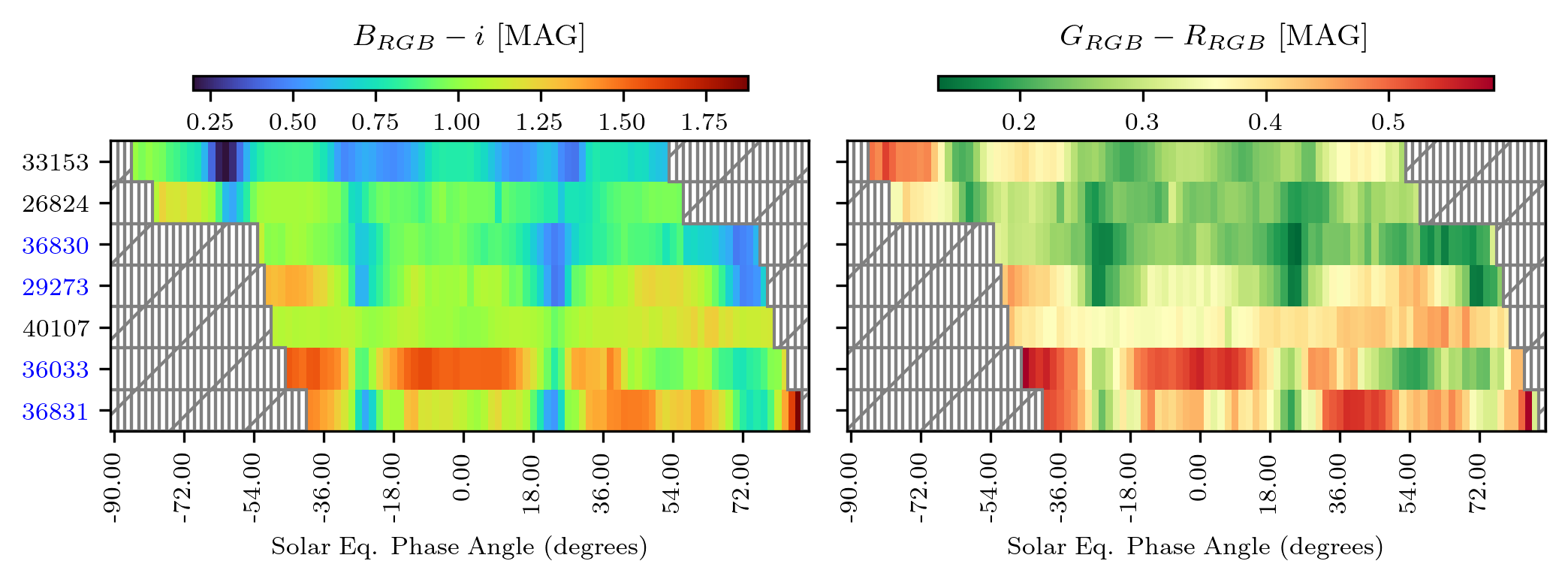}
    \caption{Colour light curves transformed into 2D image data for $B_{RGB} - i$ and $G_{RGB} - R_{RGB}$ colour indices with respect to the solar equatorial phase angle. Each row represents the colour light curve of a satellite object identified by its given NORAD ID. Blue ID labels indicate Spacebus-4000 satellites; black ID labels denote SSL-1300 satellites. Each colour light curve is split into 100 bins of equal phase width. We can note the common structures appearing within the colour light curves of all objects with three strong features appearing with two blue features occurring at  $\pm$ 25\,$^\circ$. The objects are ordered by increasing longitude relative to La Palma.}
    \label{fig:Col_ims}
\end{figure*}
\begin{table}[H]
    \centering
    \begin{tabular}{ccccccc}
    %\hline
        Night & NORAD  & $c_1$ & $\delta_1$  & $c_2$ & $\delta_2$ \\ 
        Observed & ID & (M) & (M) & (M) & (M) \\ \hline
        2023-05-13 & \textcolor{blue}{36033} & -0.64 & 0.03  & -0.59 & 0.03 \\ 
        2023-05-13 & \textcolor{blue}{36830} & -0.3 & 0.02 & -0.4 & 0.01 \\ 
        2023-05-18 & \textcolor{blue}{36831} & -0.66 & 0.02 & -0.74 & 0.03 \\ 
        2023-05-13 & \textcolor{blue}{29273} & -0.57 & 0.04 & -0.62 & 0.03 \\ 
        2023-05-13 & 40107 & -0.06 & 0.01 & -0.14 & 0.02 \\ 
        2023-04-23 & 33153 & -0.21 & 0.01 & -0.19 & 0.02 \\ 
        2023-04-25 & 26824 & -0.28 & 0.04 & -0.14 & 0.01 \\ \hline
    \end{tabular}
    \caption{Measurements and standard errors for the near-symmetrical colour changes measured in units of magnitudes (M). These features occurred for 7 satellites, each denoted here by their allocated NORAD ID. Blue ID labels indicate Spacebus-4000 satellites; black ID labels denote SSL-1300 satellites. The notation of $c$ refers to the measurements of the colour index change of $B_{RGB} - i$ and the respective errors on the first and second colour change are denoted by $\delta_{1,2}$}
    \label{tab:CC_measured}
\end{table}
We can also report the appearance of blue colour features occurring for 7 satellites near symmetrically about 0\,$^\circ$ of solar equatorial phase angle. Four of these satellites belong to the Spacebus-4000 class and 3 belong to SSL-1300 class.

As shown in Figure \ref{fig:Col_ims}, there are noticeable changes in the colour index as the night progressed for all 7 objects. In particular, the colour excursions are more visible in the $B_{RGB} -i$ colour space as opposed to the $G_{RGB} - R_{RGB}$ colour space where the excursions are not as large. We also often see a strong colour excursion either at the start of the night or at the end of the night, this feature may relate to the findings of \cite{2021spde.confE..49K} in which they see a strong brightness increases late in the observation window. They believed that this is caused by reflections off the side panel of the satellite which are very reflective.

Two of these changes of interest occur near-symmetrically at $\approx$\,$\pm$ 25\,$^\circ$ solar equatorial phase angle. We can clearly identify that the colour excursions are occurring at approximately the same equatorial phase angle for each object and there is a shift towards `bluer' colours. To constrain these features further, we calculated fitted measurements and standard errors on these 14 $RGB_{B} - i$ colour excursions.
\begin{figure*}[ht!]
    \centering
    \includegraphics{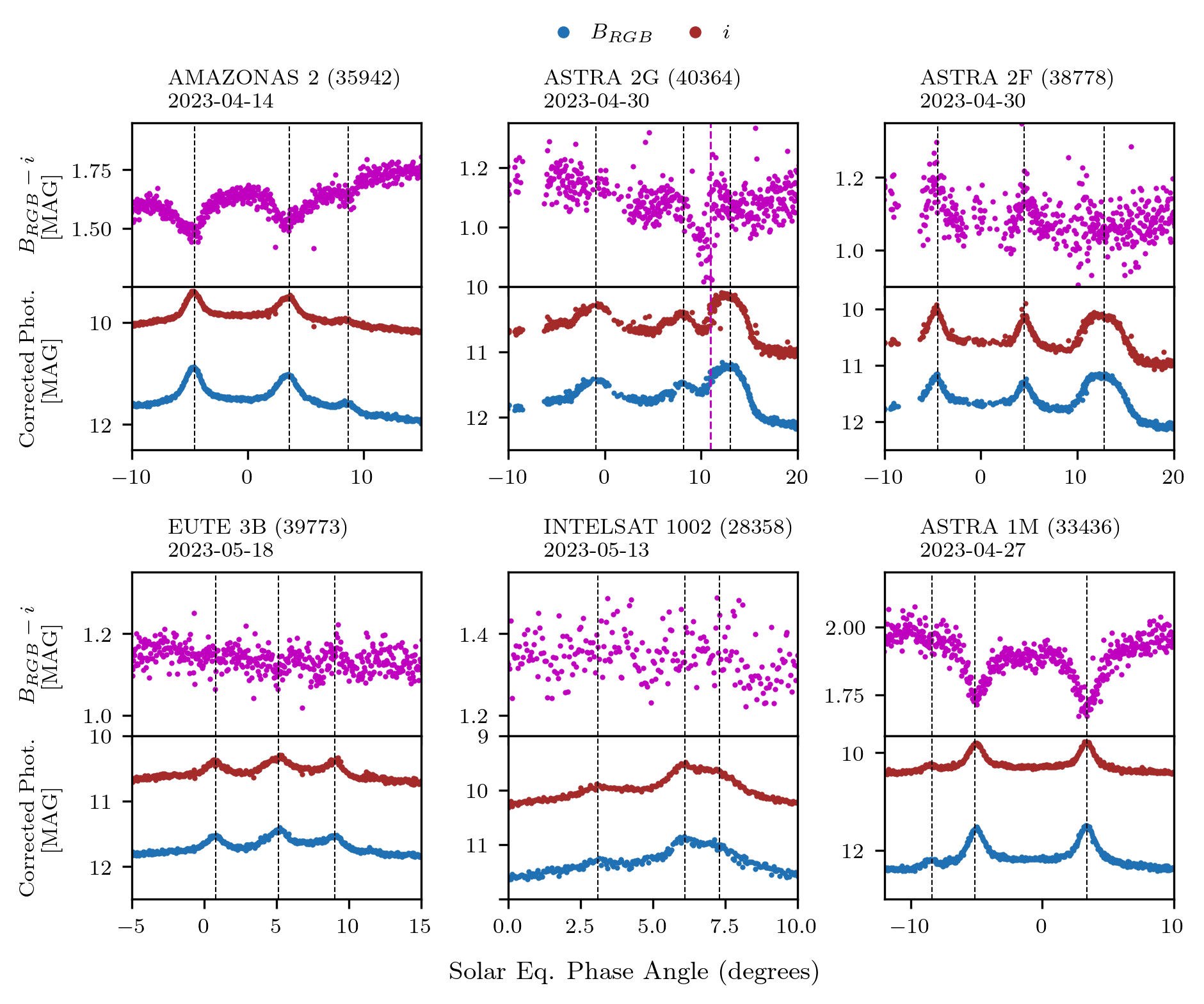}
    \caption{Triple peak features shown found within the light curves of 6 example satellites in our sample. The $B_{RGB} - i$ measurements are shown above the $B_{RGB}$ band measurements for each satellite. All satellites belong to the Eurostar-3000 bus configuration class. The larger plateau feature in the INTELSAT 10-02 single band light curve is likely due to the addition of the MEV-2 (Mission Extension Vehicle) solar panels. Black dashed lines are given at each of the glint locations and the magenta dashed line is given where we see a significant 'bluer' feature for ASTRA 2G.}
    \label{fig:triplepeaks}
\end{figure*}

To perform this calculation, we fitted across three sub-regions of the colour excursion of interest. This fitting comprised of two linear fits either side of the feature as a proxy for the baseline and a 4th order polynomial fitted across the colour change itself. The value for the colour change is approximated by taking the averages of the midpoint values of the two linear fits away from the minima of the polynomial fit, this was the most ideal way of approximating an amplitude measurement from the baseline. With this in mind, it was deemed appropriate to apply a bootstrapping technique \citep{Ader2008}. The data was re-sampled with replacement and the fitting process and colour excursions are measured with the methods described above for over 3000 iterations. The mean and standard deviation of this distribution of measured colour changes can be found in Table \ref{tab:CC_measured}.
We find that whilst indeed it is true we have limited sample size, it does appear that on average the colour excursions seen with the Spacebus-4000 class are larger than the measurements for the SSL-1300 satellites. We also see that the absolute difference between the two excursions are minimal for most satellites with Intelsat 901 (NORAD: 26824) having the largest difference between the two colour changes.

Overall, this relates to findings of \cite{doi:10.2514/1.A33583} where they find that a considerable portion of the solar spectrum greater than 600\,nm is not reflected and from that they conclude that the surfaces responsible for these glints are reflecting primarily at shorter wavelengths and absorbing photons at longer wavelengths, which is typical of older satellite solar panels. Although they are working with slitless spectroscopy and ourselves in optical photometry, we can draw the same conclusion with our findings above, where the near-symmetrical colour changes are strongly reflecting at blue wavelengths.
\subsubsection{Triple Peak Features}
We can report on some shared features that we found within certain light curves. Most notably the appearance of triple peaks, which appears to be from dual offset solar panels plus an additional specular reflection from a component. Figure \ref{fig:triplepeaks} displays clear examples of these features. 

\begin{figure}[H]
    \centering 
    \includegraphics[width = 0.5\textwidth]{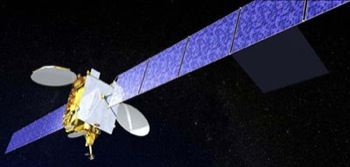}
    \caption{Artistic impression of the BADR-4 satellite. We note the `tab' like component mounted upon the solar panel as notable feature. This feature is noted to be within the designs of many Eurostar-3000 satellites as well as Eurostar-2000 \citep{Gunther-SpacePage-BADR4}.}
    \label{fig:e2000+_example}
\end{figure}

Interestingly, these satellites share a common feature found in their designs and that is a `tab' component mounted upon the solar panel, similar to the artistic impression of BADR-4 in Figure \ref{fig:e2000+_example}. The additional reflection may be caused by this. In terms of the colour response in $B_{RGB} - i$, we find that they vary across the satellites; AMAZONAS 2 shows a shift towards `bluer' colours across all three peaks, ASTRA 1M doesn't show any colour change across the smaller additional peak but does across the others, EUTE 3B and Intelsat 10-02 do not show any colour changes and ASTRA 2G shows a stepped increase in the colour index across the 2nd and 3rd peaks.

\begin{figure}[h!]
    \centering 
    \includegraphics{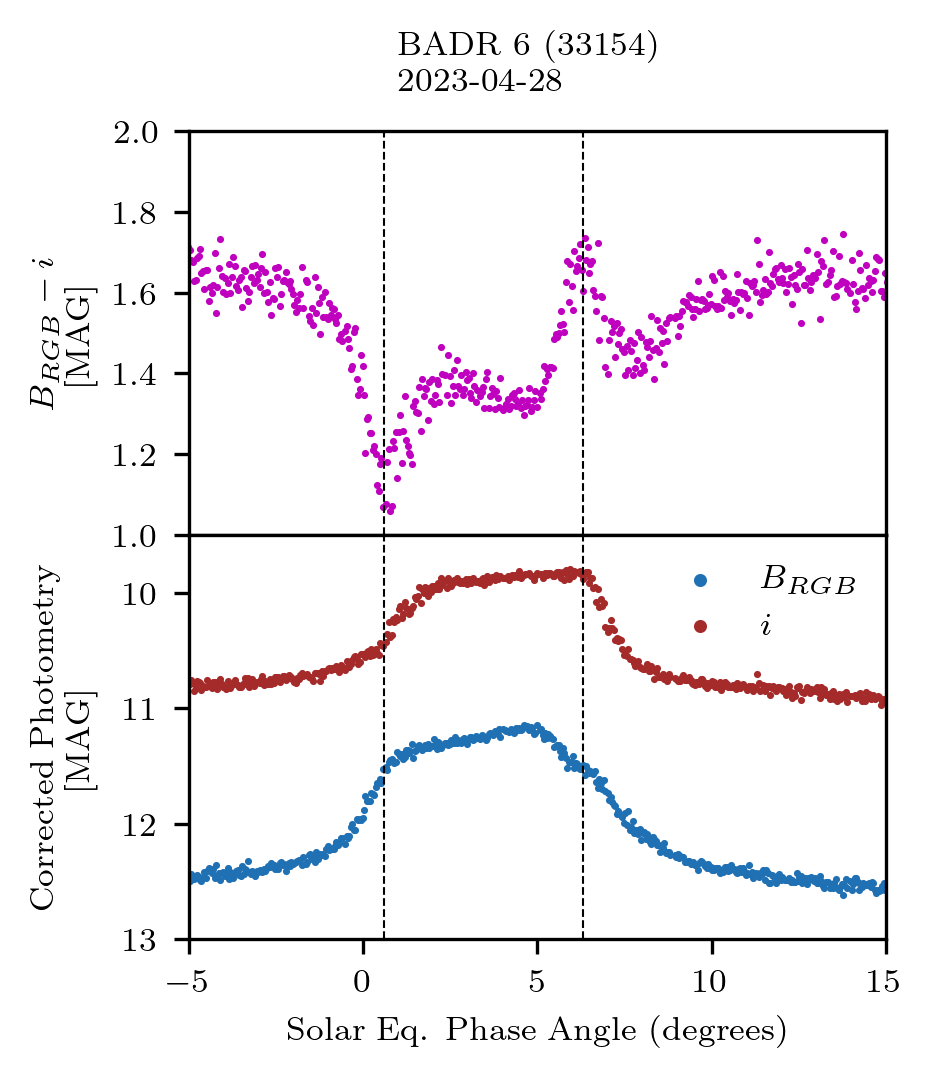}
    \caption{Portion of the BADR-6 light curve in the $B_{RGB}$ and $i$ bands. This corresponds to both and blue and then red colour responses. Black dashed lines indicate the positions of the glints seen in the colour measurements.}
    \label{fig:col_feature_}
\end{figure}

A unique colour feature is also seen within the light curve of the BADR-6 satellite. The bus configuration of BADR-6 is EUROSTAR-2000+. Figure \ref{fig:e2000+_example} displays the artistic impression of BADR-4, which is also built upon the Eurostar-2000 design. 
Figure \ref{fig:col_feature_} displays the portion of the BADR-6 light curve where in which the colour behaviour occurs. We see that as the brightness is increasing, we get first a shift towards bluer colours and then a shift towards redder colours at the end of this feature.
 The signature in the $B_{RGB}$ and $i$ bands is similar to the third peak found in the ASTRA 2F and ASTRA 2G light curves in terms of profile. The colour features are similar to that seen in the colour light curve of ASTRA 2G which is denoted by the dashed magenta line in Figure \ref{fig:triplepeaks}. It could be that the mounted solar panel component is responsible for this colour behaviour alongside favourable illumination geometry. Further observations would be able to display how this colour feature changes seasonally and constrain its geometry.

\subsubsection{Short Timescale Glints}\label{sec:glinting}
\begin{figure*}[p!]
    \centering    
    \includegraphics{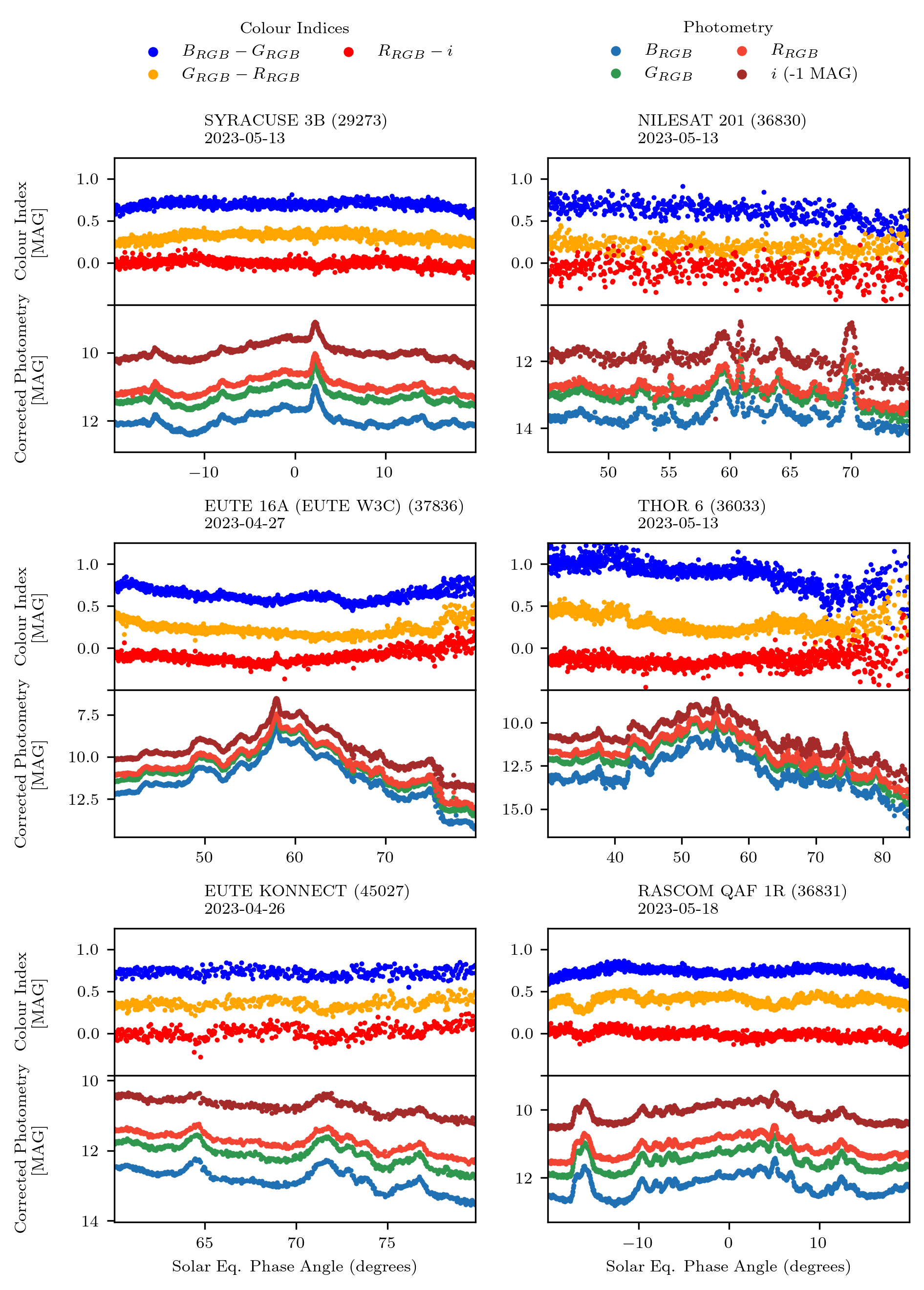}
    \caption{Examples of the glinting regions seen within light curves of satellites in our dataset. 1st row : corrected photometric measurements as a function of solar equatorial phase angle in the four colour bands. 2nd row: binned $B_{RGB} - i$ colour index as a function of solar equatorial phase angle. 3rd row: binned $G_{RGB} - R_{RGB}$ colour index as a function of solar equatorial phase angle. A 1 mag offset is applied to the $i$ band measurements.}
    \label{fig:glint_eg}
\end{figure*}
Given our observations were carried out in the weeks following the `glint season', we expected to see features with variable brightness on short timescales due to the favourable sun-observer-satellite geometry.
\begin{figure}[h!]
    \centering
    \includegraphics{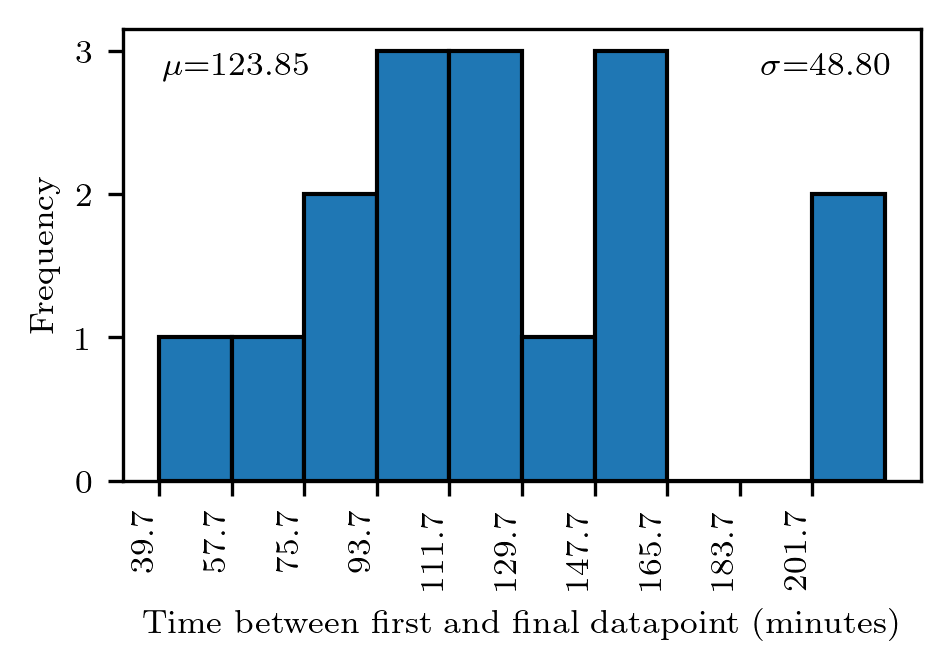}
    \caption{Distribution of the duration of glinting regions in minutes for all satellites that had glinting regions within their light curves. The mean and standard deviation of the distribution are given on the axes as $\mu$\ and $\sigma$ respectively.}
    \label{fig:glint_times}
\end{figure}
Figure \ref{fig:glint_eg} displays 6 illustrative examples of the glinting seen within 16 satellites in our sample and how these glinting features appear in colour space. The three colour indices ($B_{RGB}-G_{RGB}$, $G_{RGB} - R_{RGB}$ and $R_{RGB} - i$) were chosen to display the general colour variation seen across regions of short timescale glinting across the different wavelength bands. These 6 satellites were chosen as examples as they best highlight the uniqueness in profile. We find that the amplitudes of these glints tend to only increase by a few tenths of a magnitude and each individual glint tends to last only a fraction of a degree of phase (< 4 minutes), whilst overall the glinting regions tend to last 10's of degrees of phase. In the $B_{RGB} - G_{RGB}$ , $G_{RGB} - R_{RGB}$ and $R_{RGB} - i$ colour spaces, we see that the colour indices tend to stay fairly consistent across the glinting regions with only faint colour responses. In particular we see that there isn't necessarily a strong colour response associated with a glint or set of glints (e.g THOR-6 (NORAD: 36033), RASCOM QAF-1R (NORAD: 36831), EUTE 16A (NORAD: 37836)). 
Over the four colour bands, structurally the glinting regions have unique profiles which we would expect given the uniqueness of every satellite-sun-observer geometry and potentially the individual pointing of the satellites themselves. 
\begin{figure}[h!]
    \centering       \includegraphics{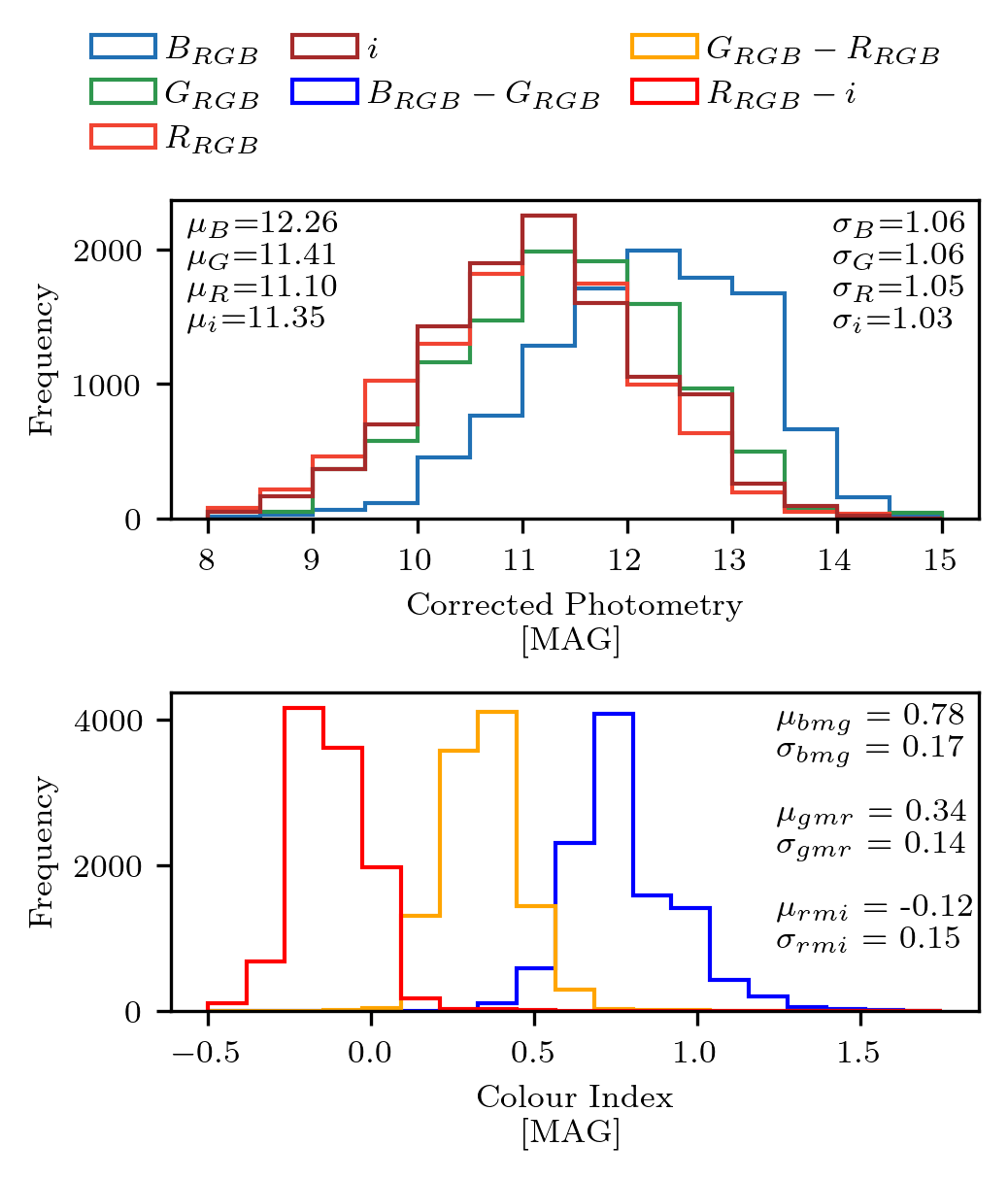}
    \caption{Photometric distributions of glinting regions within all four colour bands and the three colour index spaces ($B_{RGB}-G_{RGB}$, $G_{RGB} - R_{RGB}$ and $R_{RGB} - i$) respectively. The mean value and standard deviation are given for each distribution as $\mu$\ and $\sigma$ respectively.}
    \label{fig:glint_dist}
\end{figure}
From our measurements displayed in Figure \ref{fig:glint_times} and Table \ref{tab:glint_info}, the mean glinting time is $\approx$ 124 minutes. Figure \ref{fig:glint_dist} displays the distributions of brightness in all four colour bands and the two colour indexes considering all objects that had regions of glinting within their light curves. 
In terms of brightness we find that they tend to reach around 10th to 12th magnitude on average but can reach up to 8th in the $B_{RGB}$, $G_{RGB}$, $R_{RGB}$ and $i$ bands. 
The distributions in colour index space have small deviations especially within the $G_{RGB} - R_{RGB}$ distribution ($\sigma$ = 0.11) which informs us that generally these glints tends to be achromatic and do not generate a significant colour response which suggests that the materials of the components/mounted widgets which are responsible for these features have spectrally flat profiles. For RASCOM QAF-1R (NORAD: 36831) and SYRACUSE-3B (NORAD: 29273), which are two satellites which display the blue colour features, the glinting region is near symmetrical about 0\,$^\circ$ of solar equatorial phase angle and occurs between the two colour changes we discuss in Section \ref{sec:7CC} which suggest that the glinting is caused by off angle reflections off the same component or one located close by. 
We generally see that the colour of the glints on average is `bluer' (reflecting at shorter wavelengths) than before and after the glinting window.

\begin{table}[h!]
    \centering
    \begin{tabular}{cccc}
        NORAD & Start Time & Duration \\
        ID & (UTC) & (Minutes) \\ \hline
        42934 & 2023-04-15 03:47:17 & 119.9 \\
        29273 & 2023-05-13 22:49:20 & 160.1 \\ 
        41866 & 2023-04-13 00:48:58 & 119.9 \\ 
        39168 & 2023-04-16 02:29:21 & 79.9 \\ 
        39172 & 2023-04-23 04:13:52 & 59.9 \\ 
        28945 & 2023-04-23 21:25:32 & 99.9 \\ 
        45027 & 2023-04-27 03:14:23 & 79.4 \\ 
        44457 & 2023-04-26 20:43:05 & 109.5 \\
        37836 & 2023-04-28 01:14:50 & 159.9 \\
        41029 & 2023-04-28 20:50:36 & 39.7 \\
        36592 & 2023-04-28 23:30:19 & 99.9 \\
        40258 & 2023-05-12 21:48:48 & 139.7 \\
        36830 & 2023-05-14 03:17:46 & 119.7 \\
        36033 & 2023-05-14 01:49:44 & 214.6 \\ 
        36831 & 2023-05-18 22:12:28 & 160.1 \\ 
        39163 & 2023-05-19 01:14:12 & 219.7 \\ \hline
    \end{tabular}
    \caption{Time measurements of the glinting regions for satellites. Displayed are the measurements of the starting time (UTC) and the duration (minutes) of the glinting regions. We were careful to ensure that all glinting regions were the result of real reflections as opposed to bad zero-point measurements.}
    \label{tab:glint_info}
\end{table}

\subsection{Average Colour Measurements} 
\begin{figure*}[ht]
    \centering
    \includegraphics[width = \textwidth]{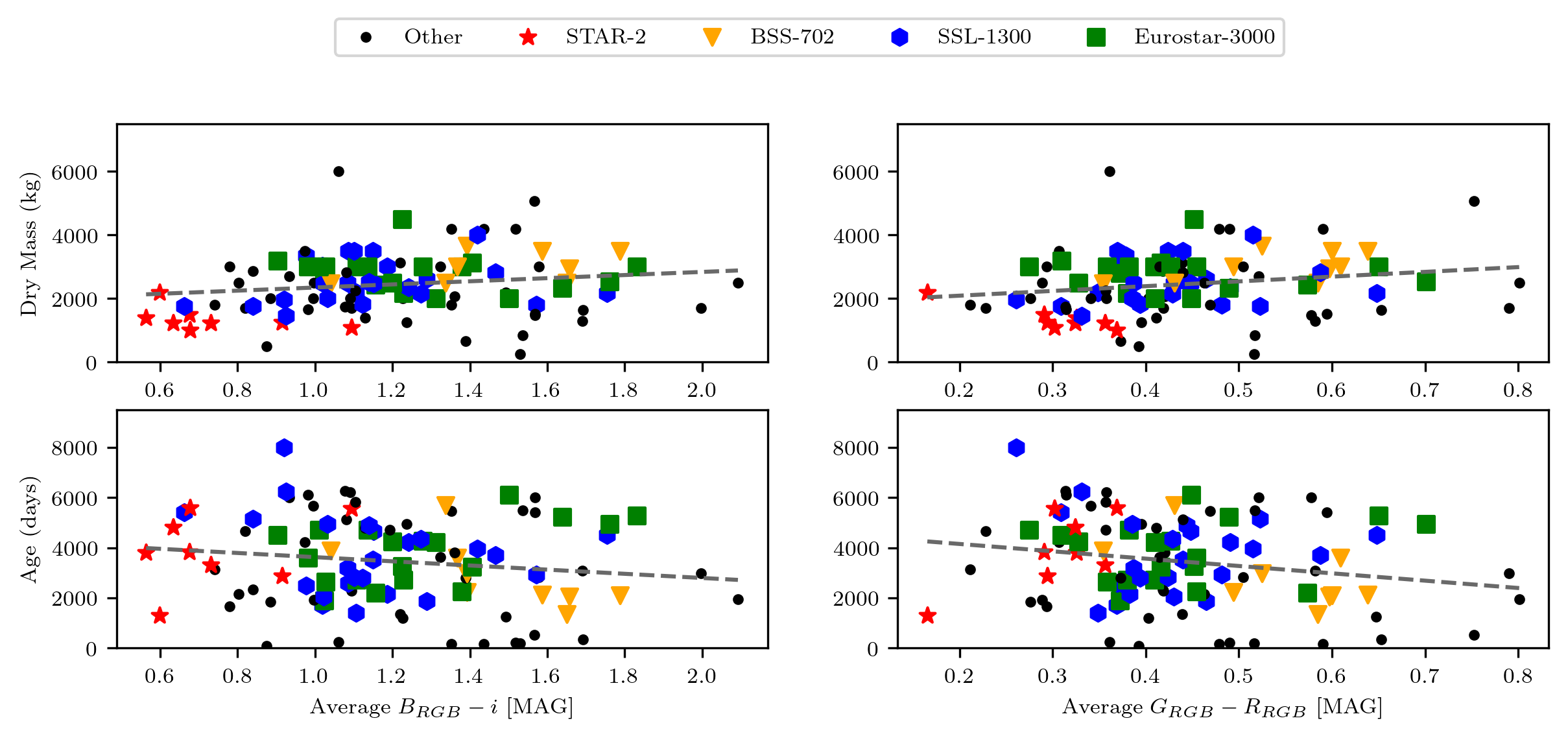}
    \caption{Time spent on orbit and dry mass for satellites with a variety of different bus configurations as a function of the average colour ($B_{RGB} - i$ and $G_{RGB} - R_{RGB}$) across their entire light curve. The grey dashed lines are linear fits applied across each population of datapoints.}
    \label{fig:Colour_rel}
\end{figure*}
Figure \ref{fig:Colour_rel} displays the average colour measurements ($B_{RGB} - i$ and $G_{RGB} - R_{RGB}$) versus the dry masses of satellites and the time spent on orbit (denoted as age in the figure). This is highlighted for satellites belonging to the most populous bus configurations.

Firstly, we find no strong overall correlation between the colour and time spent on orbit, which is in agreement with the findings of \cite{SCHMITT2020326}. We measure Pearson r correlation coefficient between colour ($B_{RGB} - i$, $G_{RGB} - R_{RGB}$) and the time spent on orbit as -0.15 and -0.21 respectively. These values suggest very weak negative correlations. The corresponding p-values are 0.13 for $B_{RGB} - i$ and 0.03 for $G_{RGB} - R_{RGB}$. These results suggest that while the correlation between $B_{RGB} - i$ and time on orbit is not statistically significant, the weak negative correlation between $G_{RGB} - R_{RGB}$ and time on orbit does reach statistical significance (p < 0.05). This implies that although both relationships are weak, we have stronger evidence for a slight tendency of $G_{RGB} - R_{RGB}$ to decrease with increased time in orbit,

Likewise, we find that the Eurostar-3000, follow a trend of increasing colour with time spent on orbit, this feature is also supported by our colour light curve maps of the Eurostar-3000 satellites (see Figures \ref{fig:bus_ims_1} and \ref{fig:bus_ims_2}).

Furthermore, we find that the STAR-2 satellites and to some extent BSS-702 satellites are isolated in colour. They have smaller and larger values of $G_{RGB} - R_{RGB}$ index respectively than satellites belonging to other bus configurations. With respect to the $B_{RGB} - i$ colour index, we find that there are comparable colours to both the STAR-2 and BSS-702 satellites and so they no longer become isolated in colour. Again, this finding is supported by the colour light curve maps (see Figures \ref{fig:bus_ims_1} and \ref{fig:bus_ims_2}). In terms of characterisation, it would be easier to discriminate between the STAR-2 and BSS-702 satellites.
\subsection{Colour Features in Relation with Physical Parameters}
\begin{figure}[hb!]
    \centering \includegraphics{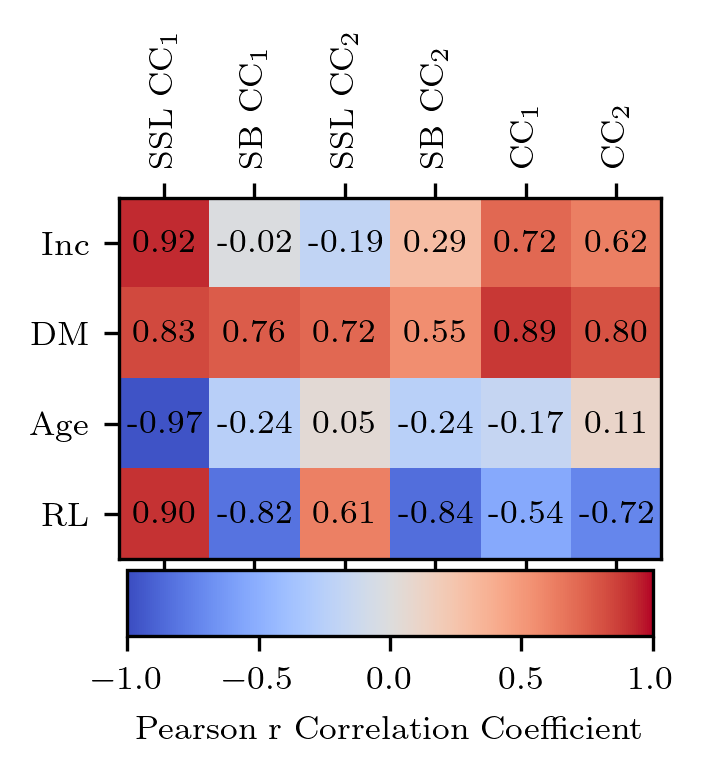}
    \caption{The measured Pearson r correlation coefficients arranged into a correlation matrix. Inc, DM and RL refer to the physical parameters of orbital inclination, dry mass and longitude relative to La Palma. SSL CC$_{1,2}$ and SB CC$_{1,2}$ refer to the 1st and 2nd colour change measurements for SSL-1300 (SSL) and Spacebus (SB) satellites respectively. CC$_{1,2}$ refers to considering the 1st and 2nd colour changes of all satellites.}
    \label{fig:Corr_Matrix}
\end{figure}
\begin{figure}[ht!]
    \centering \includegraphics{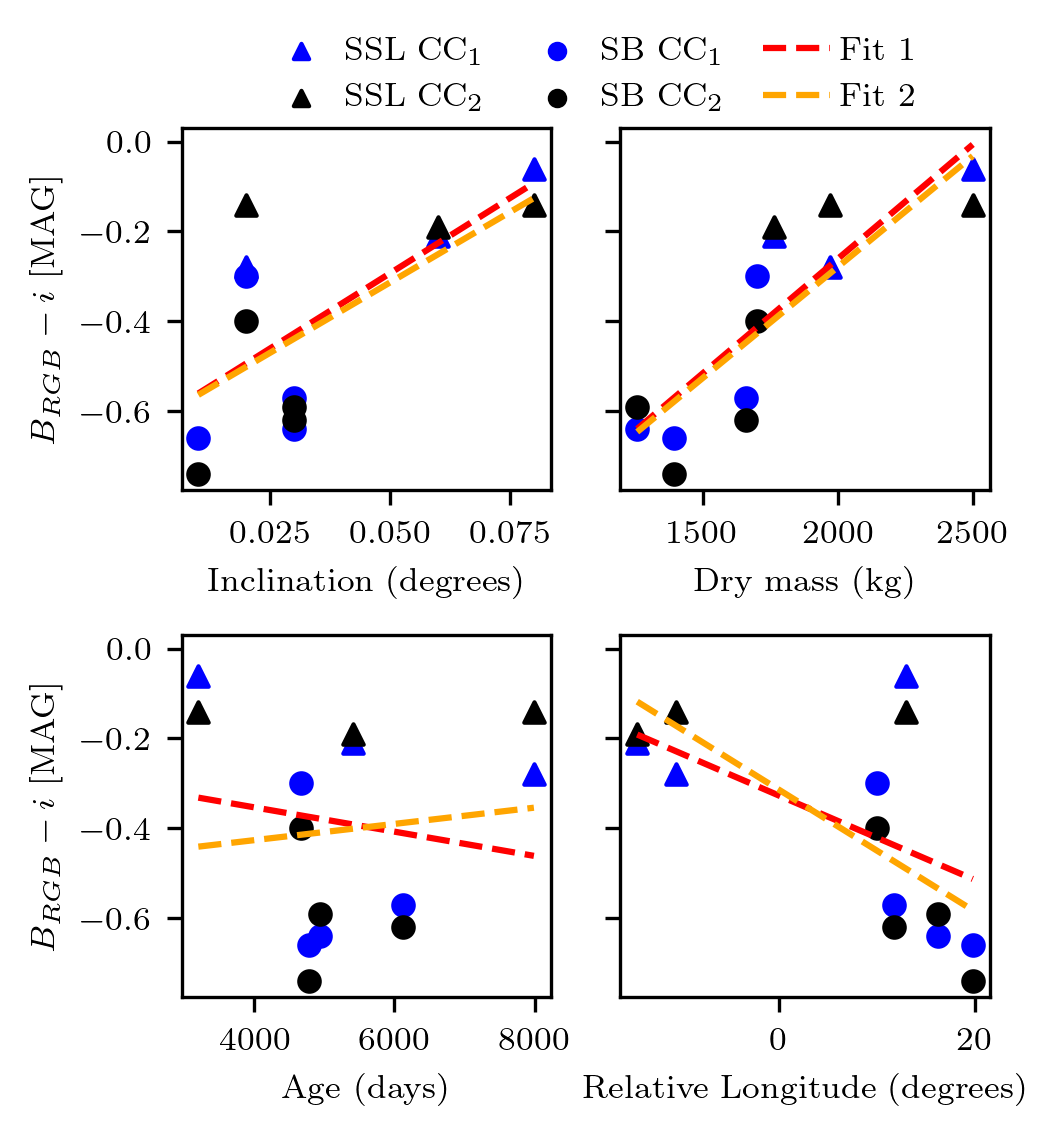}
    \caption{The measured $RGB_{B} - i$ values of the near-symmetrical colour changes as a function of different physical properties. The points in blue correspond to the first colour change (CC$_{1}$) and the points in black correspond to the second colour change (CC$_{2}$). The triangle and circle markers denote SSL-1300 (SSL) and Spacebus-4000 class satellites (SB) respectively. Linear fits are made between the 1st and 2nd colour change and the physical property, these are referred to by `Fit 1' and `Fit 2'. The errors on the colour measurements were significantly small so are not shown here.}
    \label{fig:param_plot}
\end{figure}

We also wanted to identify whether the aforementioned blue colour features in $RGB_{B} - i$ correlate with certain physical properties such as the longitude relative to La Palma, the orbital inclination, the dry mass and the time spent on orbit. The measured Pearson r correlation coefficients being outlined within Figure \ref{fig:Corr_Matrix}, with the colour changes as functions of the physical parameters are displayed in Figure \ref{fig:param_plot}.
The overall (not discriminating between SSL-1300 and Spacebus satellites) measured Pearson r correlation coefficients between each of the colour changes and orbital inclination are 0.72 and 0.62 respectively. These values suggest a strong positive correlation. However, when considering each of the bus classes separately, we measure the correlation coefficients for the 1st and 2nd colour change to be (0.92, -0.19) and (-0.02, 0.29) for the SSL-1300 and Spacebus satellites respectively. This contradicts the overall correlation we measure and is contradictory between the same bus class. Similarly, there is a very strong positive correlation between each of the colour changes and dry mass, with measured Pearson r coefficients of 0.89 and 0.80 respectively. Considering the r correlation coefficients for the colour changes of each bus class with the dry mass, we measure agreement that there is a strong positive correlation between them. There are strong negative correlations between the colour and longitude relative to La Palma with the measured Pearson coefficients being -0.54 and -0.72 respectively. This suggests that the further east the satellites are from La Palma, the `bluer', the colour changes are. Considering each bus class, the Spacebus satellites follow the overall correlation and the SSL-1300 satellites follow the inverse, becoming `bluer', the further west they are from La Palma. However, the SSL-1300 satellites are biased towards further western longitudes than La Palma.

However, there are very weak correlations between each of the colour changes and age with measured Pearson coefficients of -0.17 and 0.11 respectively. The correlation coefficients for the Spacebus satellites and the second colour change of the SSL-1300 satellites are in agreement with this overall correlation. However, we do measure the first colour change of the SSL-1300 satellites to have a very strong negative correlation with age (r = -0.97). The fitting in Figure \ref{fig:param_plot} between the colour changes and age is also poor, which further suggest that there is a lack of an overall relationship. We note that there are only 14 datapoints (2 colour change measurements for each satellite) overall with fewer when considering either the Spacebus-4000 or SSL-1300 satellites individually, which is a small sample. Similarly, there are also a lack of datapoints across parameter space e.g no datapoints between -5\,$^\circ$  and 5\,$^\circ$ of relative longitude. More measurements and longer term measurements would be needed to probe whether a relationship does exist between the size of these colour changes and each of the physical parameters outlined.

\subsection{Solar Panel Offsets}\label{sec:SP_OFF}
\begin{figure}[h!]
    \centering
    \includegraphics{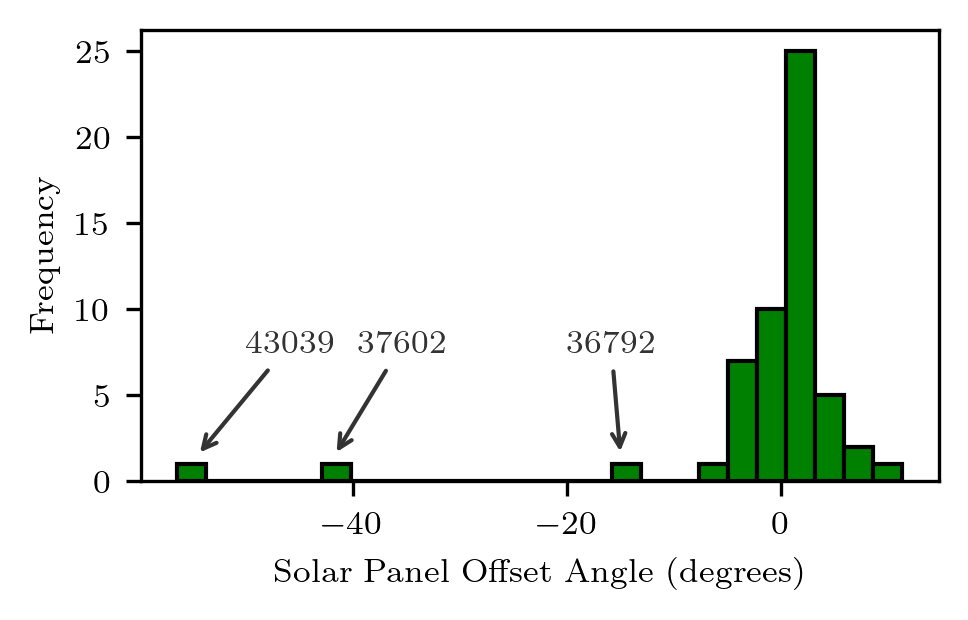}
    \caption{The distribution of solar panel offsets for 54 satellites in our sample. For satellites that have solar panels which are offset from each other, we take the average between the two offsets. The three satellites that lie away from the main distribution are annotated by their NORAD IDs.}
    \label{fig:offsets_dist}
\end{figure}

\begin{figure*}[t!]
    \centering
    \includegraphics{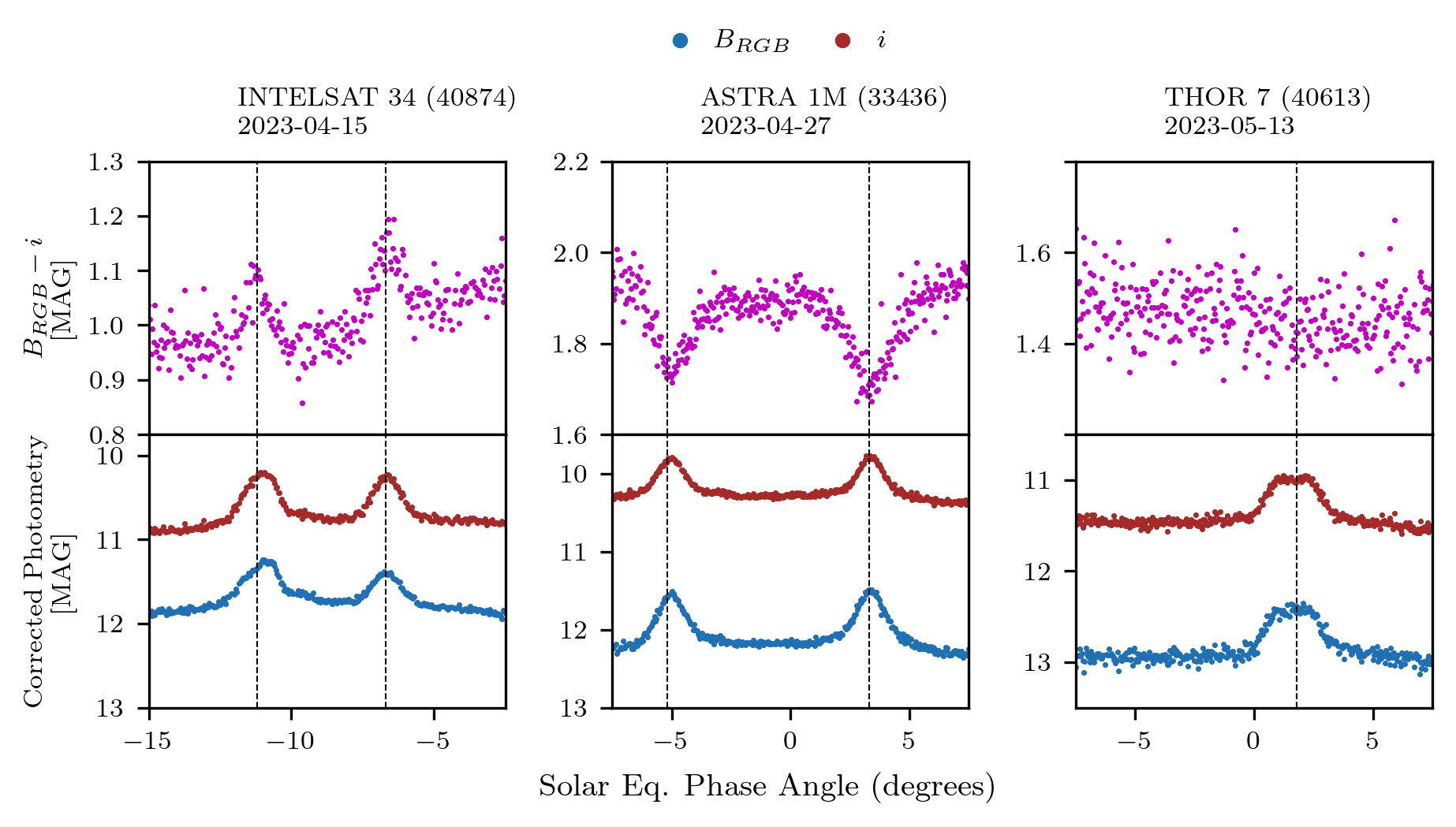}
    \caption{Three examples of the colour response across glints caused from specular reflections off of the solar panels. Upper: Colour measurements in the $B_{RGB} - i$ band. Lower: measurements made in the $B_{RGB}$ (blue) and $i$ (brown) bands. From left to right, we note a shift towards redder colours, towards bluer colours and no significant colour change. Black dashed lines are used to highlight the locations of the solar panel glint(s) in both the single band and colour measurements.}
    \label{fig:example_sp_glints}
\end{figure*}

Solar panel offsets can be determined directly by measuring the offset of specular glints. Assuming that the glints correspond to a reflection from a flat surface, then the normal must be in the direction of the phase angle bisector (PAB). We measure the offset of the specular glints from 0\,$^\circ$ phase angle and take half of this value to yield the solar panel offset \citep{Payne2006SSAAO}. Using this methodology, we are able to measure solar panel offsets for 54 satellites, 26 of these satellites have both panels which are offset at slightly different angles which presents itself in the light curve with m-shaped signatures (see Figure \ref{fig:example_sp_glints}). The distribution of these measurements are shown in Figure \ref{fig:offsets_dist}. We find that most of these offsets are between -10 and 10\,$^\circ$, which is typically what we would expect for geostationary satellites \citep{Payne2006SSAAO}.

There are three outliers within this distribution. Firstly, satellite Telstar 14R (NORAD: 37602) which has an offset of approximately -41.5\,$^\circ$. It is understood from \cite{20210020397} that the northern solar array failed to fully deploy diminishing the amount of power available for transponders and thus reduced the life expectancy of the satellite.

Secondly, we note that ALCOMSAT-1 (NORAD : 43039) has two solar panels offset at very large angles to what we normally expect, approximately -57.8 and -54.9\,$^\circ$ respectively. The satellite has spent almost six years on orbit so it is unlikely that this large offset is due to management of the solar arrays which is typically done in the early time spent in orbit. There is a possible link that these large offsets are implemented due to the bus on which it is configured. ALCOMSAT-1 is configured with the Chinese DFH-4 bus in which it provides high capability in output power and communication capacity. The approximate dry mass of ALCOMSAT-1 is 2500kg \citep{JSR-SATCAT}. Considering these details, it is possible that the large offsets are to balance the impact of any torque generated by solar radiation pressure impacting upon the satellite. Finally, Echostar-15 (NORAD : 36792) which lies closer to the main distribution than the previous outliers, has an offset of about -15.7\,$^\circ$.

\subsubsection{Solar Panel Glint Colours}
\cite{2015amos.confE..74J} and \cite{2014amos.confE..31B} highlight that the illumination-observational geometry plays a significant role in the observed colour response from solar panel glints. The spectral energy density (SED) of reflected light off of a surface such as the solar panels is not invariant but varies and it is this variation of the overall geometry that leads to differences in the observer colour index of the solar panels. For example, at times around the summer and winter solstices, the angle at which the sun’s light illuminates a geostationary satellite solar array is the greatest. The specular reflection is thus directed away from Earth’s equator; to the north during the northern hemisphere winter, and to the south during the northern hemisphere summer. For example, since the illumination angle is large on these nights close to the solstices and if the observer is more northern than where the specular reflection is directed, the extra contribution made by the solar panel glint to the measured flux, above the level measured without the specular reflection, consists of more longer wavelength light than shorter wavelength light. Thus, the overall colour shifts towards `redder' colours.
\begin{table}[ht!]
    \centering
    \begin{tabular}{|c|c|c|c|c|}
    \hline
        Type & B & N & R & Total \\ \hline
        Single & 8 & 11 & 9 & 28 \\ \hline
        Dual & 6 & 13 & 7 & 26 \\ \hline
        Total & 14 & 24 & 16 & 54 \\ \hline
    \end{tabular}
    \caption{Colour responses of the measured solar panel offsets in our sample by the type of offset. B, N and R refer to the colour response of shifting towards bluer colours, no observed colour response and shifting towards redder colours respectively. Single and Dual refers to where the solar panels were both offset at the same angle or at slightly different angles respectively.}
    \label{tab:sp_colour}
\end{table}

We find three types of colour responses exhibited by the solar panel glints; stable behaviour, shifts toward redder colours and bluer colours respectively. Table \ref{tab:sp_colour} shows the qualitative measurements of the solar panel glint colours in $B_{RGB} - i$. The majority of the colour responses are that there is no significant colour change across the solar panel glints.

Figure \ref{fig:example_sp_glints} highlights an example of each of these colour responses. There are no clear correlations of the type of colour response with the orbital slot longitude, age or bus configuration. The variation of the colour of the solar panel glints could possibly be due to the illumination angle at the time at which they were observed, seasonal changes in solar declination will result in variable solar array illumination angles. That is to say that if we observed each of these solar panel glints at different points in the year we would see variable colour responses. Further observations of these satellites would be needed to verify this.

Finally, 52\,$\%$ of the solar panels that had dual offsets belonged to the Eurostar-3000 bus configuration class, which suggests this is a common operating mode.

\section{Discussion}\label{sec:discussion}

\subsection{Discussion of Colour Measurements}
The colour light curve maps introduced in this paper serve as an extremely useful way to compare and contrast satellites belonging to different bus configurations. Our key findings from these measurements are as follows:
\begin{itemize}
    \item Older Eurostar-3000 satellites are `redder' than their successors, which contradicts the trend seen within the other bus classes
    \item STAR-2 satellites as a group are much `bluer' than any other configuration and have colours comparable to the Sun
    \item BSS-702 satellites as a group are `redder' than any other configuration
\end{itemize}

The colour evolution within the Eurostar-3000 bus configuration class could point towards manufacturing changes during the time when ASTRA-3B (NORAD: 36581) was being constructed. Further observations of more Eurostar-3000 satellites are needed to confirm whether this is a definitive trend. If it is, then this would be a powerful tool for identifying specific Eurostar-3000 satellites. It is also a possibility that the similarity in colour between the five red satellites are linked with their operators requirements at the time, for example three of these satellites are operated by EUTELSAT and are identical in design, these being Eute 33E (NORAD: 33750), Hotbird 13B (NORAD: 29270) and Hotbird 13C (NORAD: 33459). There is also evidence to support that this might be the result of manufacturing changes given the project documentation for which the Eurostar-3000 mechanical platform would be improved upon in \cite{ESA-CSC} and it is mentioned that solar panel manufacturing received two new design processes. The status update of this document was given in 2014, which would be before the majority of the satellites in our Eurostar-3000 sample were launched. We don't see any evidence to suggest that the evolution of colour is related to either the satellites on sky position or pointing given multiple satellites share similar communication footprints and have different colours. There is also no correlation with the solar declination angle.

\begin{figure}[H]
    \centering
    \includegraphics[width = 0.5\textwidth]{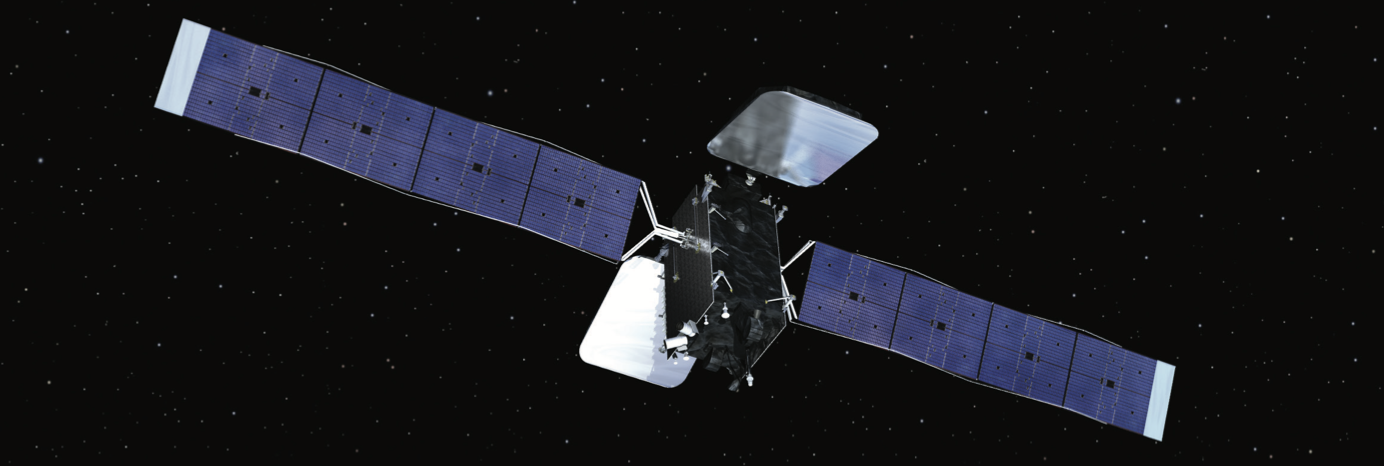}
    \caption{An artistic impression of the STAR-2 bus configuration design, we see the most notable feature of this design are the large flat antenna reflectors \citep{GEOSTAR-2-Datasheet}.}
    \label{fig:STAR2-eg}
\end{figure}

Furthermore, STAR-2 satellites have a typical design with large flat antenna reflectors as shown in Figure \ref{fig:STAR2-eg}, these are typically made from aluminium which as a material has a greater spectral reflectance value at shorter wavelengths \citep{article}. This could explain why we see these satellites being much `bluer' than any other configuration overall. It could also be a possibility that the STAR-2 satellites have less gold MLI materials either within their designs than the BSS-702 satellites or in the line of sight of STING.

With regards to the blue colour features seen within the SSL-1300 and Spacebus-4000B class satellites, we believe that these features could be caused by reflections of inclined components mounted upon the solar panels given their near-symmetrical nature and `blueness' compared to the rest of the colour light curve. We see some near-symmetry with regards to the glinting features within RASCOM QAF-1R (NORAD: 36831) and SYRACUSE-3B (NORAD: 29273) which would further support this argument. The older SSL-1300 satellites show these near-symmetrical colour changes, which suggests that the newer satellites within this class have moved away from designs with the component which is causing this. We plan to follow up with further observations of satellites displaying these features with the aim of explaining the origin and to outline any relationship with the solar declination angle.

\subsection{Discussion of Short Timescale Features}

The short timescale glinting features we find within our GEO-satellite sample are unique to each satellite and we think are a result of reflections off flat surfaces such as solar panels. This is the attribution given to the THOR-6 glints within the works of \cite{WIERSEMA20223003}. 

The glinting features tend to not generate a large colour response over an average duration of two hours. However, we do find that the average colour of the glinting regions is `bluer' compared to before and after these regions. If these unique features do appear seasonally, this would be a very powerful characterisation tool for identifying individual objects.

\section{Conclusions}\label{sec:summary}
Full night multi-colour light curves for over 100 GEO satellites were obtained using the STING instrument following observations over five weeks. This was the first ever large systematic survey of the GEO regime from which we present the analysis of full night light curves from our observations. We characterised the noise performance of STING and quantified the error reduction between simultaneous and sequential observations.

We were able to resolve and detect manoeuvres in the light curve of the LDPE-3A satellite. Colour measurements were transformed into colour light curve maps which gave a relative comparison between satellites belonging to different bus configuration families. From this and further colour analysis, we found that the STAR-2 and BSS-702 satellites are the most easily identified classes due to their colours and that there is tentative evidence showing that older Eurostar-3000 satellites are redder. 

We also investigated regions of short timescale glints across multiple satellites and found that they are unique in profile to each satellite and imply possible reflections of inclined flat surfaces such as solar panels given their relative blueness with respect to the rest of the colour light curve, this uniqueness could serve as a very powerful characterisation tool for identifying individual satellites, especially if these profiles are seasonal. We investigated the changes in colour of the eastern and western illuminated faces of satellites and found the Spacebus-4000 satellites display the largest changes in colour collectively than any other class.

Finally, we identified 54 solar panels offsets in our sample and investigated the colour response of these solar panel glints. We found that firstly, we see variation in the type of colour response and secondly, the majority did not have any associated colour change change across the glints compared to them shifting towards `redder' or `bluer' colours.

Future work will focus on the analysis of retired and tumbling satellites observed within this survey. There will also be a focus on further observations of the satellites for which we found the blue colour changes from both the eastern and western face of the light curve to determine the origin and relative geometry of these features.

\section{Acknowledgements}
This work makes use of data from the STING instrument operated on the island of La Palma by the University of Warwick in the Spanish Observatory del Roque de los Muchachos of the Instituto de Astrofísica de Canarias.

JAB acknowledges support from the Science and Technology Facilities Council
(grant ST/Y50998X/1).
For the purpose of open access, the author has applied a Creative Commons Attribution (CC-BY) licence to any Author Accepted Manuscript version arising from this submission.

{
\section*{Data availability}
The data underlying this article will be shared on reasonable request to \texttt{P.Chote@warwick.ac.uk}.
}

\appendix
\section{Noise Calculations}\label{appendix:noise}

The various contributory sources of our theoretical noise model were calculated using Equations \ref{eq:ps_noise} - \ref{eq:tot_noise} for each of the four filter bandpasses.

$f_{target}$ refers to the synthetic target flux in ADU captured within the entire aperture and $f_{sky}$ refers to the average calculated sky background value which was derived from a date where the moon disk was most illuminated as it passed through the meridian.

$N_{Photon}$, $N_{Sky}$, $N_{DC}$, $N_{R}$, $N_{S}$ and $N_{T}$ refer to the photon shot noise, sky background noise, dark current noise, read noise, scintillation noise and total noise respectively. The modified Young approximation used to calculate the scintillation noise, i.e., Equation \ref{eq:scint_noise} and consequently the empirical coefficient $C_Y$ are taken from \cite{Osborn_2015}.

\begin{equation}
    N_{Photon} = \frac{\sqrt{f_{target}}}{f_{target}}
    \label{eq:ps_noise}
\end{equation}

\begin{equation}
    N_{Sky} = \frac{\sqrt{f_{sky}}}{f_{target}}
    \label{eq:sky_noise}
\end{equation}

\begin{equation}
    N_{DC} = \frac{\sqrt{DC}}{f_{target}}
    \label{eq:dc_noise}
\end{equation}

\begin{equation}
    N_{R} = (\sigma_{r}^{2}*n_{pix})
    \label{eq:read_noise}
\end{equation}

\begin{equation}
    N_{Scint} = f_{target}*\sqrt{10^{-5}*C_{Y}^{2}*D^{-\frac{4}{3}}*t^{-1}*(AM)^{3}*e^{\frac{-2h}{h_{0}}}}
    \label{eq:scint_noise}
\end{equation}

\begin{equation}
    N_{Total} = \sqrt{f_{target} + f_{sky} + N_{R}+DC + (N_{scint})^{2}}
    \label{eq:tot_noise}
\end{equation}

%% Bibliography
%% Author year style
\bibliographystyle{jasr-model5-names}
\biboptions{authoryear}
\bibliography{refs}

\end{document}